\documentclass[final,3p,times]{elsarticle}

\usepackage{amssymb}
\usepackage{lipsum}
\usepackage{amsmath}
\usepackage{algorithm}
\usepackage{algpseudocode}
\usepackage{graphicx}
\usepackage{xcolor}
\usepackage{subcaption}
\usepackage{multirow}
\usepackage{siunitx}
\usepackage{url}
\journal{abcdefg}

\begin{document}

\begin{frontmatter}



\title{On Multifidelity Neural Operator for Data-Driven PDE Predictions}


\author[inst1,inst4]{Ghifari Adam Faza}
\author[inst2]{Kemas Zakaria}
\author[inst2]{Pramudita Satria Palar}
\author[inst1,inst4]{Keivan Shariatmadar}
\author[inst2]{Lavi Rizki Zuhal}
\author[inst3]{Hans Hallez}
\author[inst1,inst4]{David Moens}

\affiliation[inst1]{organization={Mecha(tro)nic System Dynamics (LMSD), Department of Mechanical Engineering, KU Leuven},
            city={Heverlee},
            postcode={3001},
            country={Belgium}}

\affiliation[inst2]{organization={Faculty of Mechanical and Aerospace Engineering, Institut Teknologi Bandung},
            city={Bandung},
            postcode={40132},
            country={Indonesia}}

\affiliation[inst3]{organization={Distributed and Secure Software (DistriNet), Department of Computer Science, KU Leuven},
            city={Brugge},
            postcode={8200},
            country={Belgium}}
            
\affiliation[inst4]{organization={FlandersMake@KU Leuven},
            country={Belgium}}

\begin{abstract}
Solving physical problems governed by partial differential equations (PDEs) is often time-consuming and costly. These limitations frequently restrict the number of samples that can be generated, posing challenges for accurate downstream analyses. Recently, operator learning frameworks, such as neural operators, have gained attention for their ability to learn nonlinear mappings between infinite-dimensional function spaces. However, these models require substantial amounts of high-fidelity data for training, which is prohibitively expensive to generate. To address this issue, we explore a multi-fidelity learning approach that leverages a large amount of inexpensive data sources that are well-correlated with a limited set of high-fidelity data. Additionally, we observed that many existing multi-fidelity studies rely primarily on test cases where fidelity levels differ only by grid resolution. While such scenarios are valid and practical, they often result in lower-fidelity data that closely resembles the structure of the high-fidelity counterpart. This scenario captures only a narrow subset of the broader landscape of the multi-fidelity learning problem. For instance, grid-based fidelity differences fail to capture the more substantial model discrepancies seen in computational fluid dynamics (CFD), such as those between Large Eddy Simulation (LES) and Reynolds-Averaged Navier–Stokes (RANS). These types of differences present richer, more complex challenges that make multi-fidelity learning significantly more compelling. We evaluate several proposed multi-fidelity architectures across a range of PDE test cases: two cases of multi-fidelity by grid resolution, one case of a toy problem that resembles multi-fidelity by governing equation, and one case of an unsteady multi-fidelity problem. While the first two serve as benchmark cases, the remaining tests the approach's robustness in capturing more complex correlations. Our findings indicate that not all multi-fidelity neural operator architectures outperform purely high-fidelity models under limited data conditions, and we have to be careful with our multi-fidelity strategy. Among all approaches, only the transfer learning strategy consistently outperforms the high-fidelity baseline, highlighting its potential for efficient and accurate PDE solutions.
\end{abstract}


\begin{highlights}
\item Training a machine learning-based PDE solver, such as a neural operator (NO), requires a considerable amount of high-quality and high-fidelity data to generalise the solution accurately. However, access to high-fidelity data is costly and often limited. Thus, a multi-fidelity learning strategy offers a solution to address this problem.
\item There are multiple strategies and approaches for constructing a multi-fidelity learning model for a neural network-based model. This paper compares and evaluates different approaches to enable multi-fidelity learning in the neural operator model.
\item We observed that many recent studies assume that multi-fidelity data arise primarily from differences in discretisation resolution. While this is a valid and practical scenario, the resulting gap between high-fidelity and low-fidelity data is often relatively small. Consequently, we argue that the impact of incorporating lower-fidelity data in such cases may be limited and less representative of broader multi-fidelity learning challenges.
\item In this paper, we propose a modification of the current multi-fidelity test case baseline to resemble real-world cases. We also propose a more complex test case which includes a time-dependent component.
\end{highlights}


\end{frontmatter}


\section{Introduction}\label{sec:introduction}
Modelling engineering and science problems often involves solving complex partial differential equations (PDE) systems, or through experiments. However, the computational complexity and resource-intensive nature of these methods frequently result in prohibitively expensive single-configuration analyses. Such constraints severely limit the number of samples that can be generated, creating significant challenges for critical downstream tasks, including reliability assessment, design optimisation, and inverse engineering problems. Thus, faster alternatives are needed to address this problem. While hardware acceleration has improved the efficiency of numerical PDE solvers \cite{Huthwaite2014, Bernardini2021}, data-driven methods are gaining popularity as a viable alternative. These surrogate models, such as support vector regression (SVR) \cite{Drucker1996}, Gaussian processes (GP) \cite{Rasmussen2005-at}, and neural networks (NN) \cite{tripathy2018deep, sun2019review}, learn input-output relationships from existing data, providing efficient and scalable solutions.

As scientific problems grow increasingly complex, such as accurately modelling full-field PDE solutions, the need for models capable of capturing intricate spatial and temporal patterns has grown significantly. Traditional methods such as SVR and Gaussian processes GP, often paired with reduced-order modelling techniques like proper orthogonal decomposition (POD) \cite{chatterjee_pod}, have been employed to tackle such challenges. However, recently neural networks have emerged as a powerful and flexible alternative, owing to their high expressive capacity, advances in computational resources, scalable architectures, and access to large datasets. These qualities make them especially suitable for complex predictive tasks in scientific machine learning. Applications of neural networks in this domain include solving inverse problems governed by PDEs \cite{Pakravan2021, molinaro2023, Berg2021}, data-driven discovery \cite{raissi2018ddr, Pan_2018}, and design optimisation \cite{Rai2000, Du2021}. The great flexibility of NN architecture is the dominant factor that drives the superiority of neural networks. For example, classical artificial neural networks (ANNs), or multi-layer perceptrons (MLPs) \cite{Murtagh1991, Popescu2009}, use stacked layers of perceptrons with linear transformations and nonlinear activations. When dealing with complex data types such as images, the standard nonlinear layer is then replaced by convolutional layers \cite{Lecun1998}, allowing the model to more effectively capture local spatial structures.

Neural networks have long been used as surrogate models, but the introduction of physics-informed neural networks (PINNs) \cite{Raissi2019} marked a significant shift in this approach. Unlike standard neural networks, PINNs incorporate external information, such as physical laws, directly into the loss function. This integration allows the model to learn more efficiently and accurately. Notably, while PINNs modify the loss function, they do not alter the network’s architecture or layer structure. Deep operator network (DeepONet) \cite{Lu2021}, on the other hand, learns nonlinear continuous operators via two sub-networks, namely, trunk-net and branch-net. The former is used to encode the location of the output functions, and the latter is used to encode the input function at a fixed number of sensors/discretisation. More recently, the neural operator (NO) \cite{Kovachki2021} has emerged as a resolution-invariant alternative to DeepONet. At its core, NO leverages an integral operator layer, enabling it to learn mappings between infinite-dimensional function spaces, offering greater flexibility and generalisation across different resolutions.

Advancements in neural network-based models have significantly improved their ability to capture complex physical phenomena. However, these models require large volumes of high-quality training data to achieve accurate learning, leading to substantial computational and financial costs for dataset generation. Addressing this challenge necessitates a more efficient approach. One way to mitigate computational expenses is by using coarser discretisation grids or simplified governing equations to generate lower-fidelity solutions. However, these lower-fidelity simulations often lack accuracy and fail to capture detailed phenomena that their high-fidelity counterparts are able to resolve, such as small-scale turbulence in CFD simulations. Despite this limitation, lower-fidelity solutions generally preserve the overall structure of the high-fidelity solutions. Given the ease of generating large quantities of lower-fidelity data and the constraints on producing high-fidelity datasets, a natural strategy is to use lower-fidelity data to learn the general structure of PDE solutions while leveraging limited high-fidelity data to refine finer details. This approach, known as multi-fidelity learning, enables models to achieve greater predictive accuracy than single-fidelity models trained with only a small amount of high-fidelity data.

The concept of multi-fidelity learning is well-established in other machine learning techniques, such as Gaussian process regression \cite{Myers1982, Forrester2007} (co-Kriging) and polynomial chaos expansion \cite{Palar2016} with applications in design optimisation \cite{Koziel2014, Liu2022}. In the context of neural-network-based models, several approaches such as intermediate output architecture \cite{Meng2020, Guo2022}, multi-step architectures \cite{Guo2022}, and transfer learning method \cite{Chakraborty2021} have been proposed to tackle the multi-fidelity learning problems. Multi-fidelity learning has effectively been utilised in neural operators, such as multi-fidelity neural operators with transfer learning \cite{Lyu_2023,tang2024}, and multi-fidelity wavelet neural operators with residual value architecture \cite{Tripura2024}. These studies have successfully showcased the benefits of multi-fidelity learning across various test cases. However, in all the multi-fidelity neural operator test cases we found \cite{Lyu_2023, Howard2023, tang2024, Tripura2024}, fidelity variations are introduced solely through differences in grid resolution, where lower-fidelity solutions resemble higher-fidelity ones but at a coarser scale. While such cases are valid examples of multi-fidelity learning, they do not capture variations in the governing equations. For instance, distinct flow separation characteristics around an airfoil at high angles of attack \cite{renard2021}, as modelled by Reynolds-averaged Navier-Stokes (RANS) and detached eddy simulation (LES), cannot be replicated merely by adjusting grid resolution. This highlights the need for multi-fidelity approaches that account for fundamental differences in the underlying physics.

In this study, we assess various architectures and methodologies for multi-fidelity learning in neural operators. First, we perform a comparative and ablation study on different architectures using benchmark test cases, including the one-dimensional stochastic Poisson equation and the two-dimensional Darcy flow at varying resolutions. Then, to resemble multi-fidelity in terms of governing equations, we propose a modification on the existing 2-dimensional Darcy flow and perform a comparative study on different architectures and strategies. Lastly, we propose a new benchmark multi-fidelity test case on unsteady smoke inflow, which includes the time-dependent component. \textcolor{red}{Specifically, we compared several approaches for multi-fidelity NO learning, namely, the intermediate architecture, the multi-step architecture, and the transfer learning.}

The remainder of the paper is organised as follows: Section \ref{sec:mfno overview} provides an overview of the Neural Operator (NO) model and multi-fidelity learning. Section \ref{sec:experiments} introduces the benchmark test cases along with our proposed test case. Finally, Section \ref{sec:conclusion} presents the concluding remarks.

\section{Multifidelity Neural Operator}
\label{sec:mfno overview}
This section provides a brief overview of the NO model and multi-fidelity learning. The brief overview of the theoretical foundation of neural operators provides a mathematical primer to the readers on the concept of learning mappings between two infinite-dimensional spaces of functions. We also provide a quick review of multi-fidelity learning, primarily on NN-based models, to provide the reader with the general idea of learning with limited high-fidelity data.

\subsection{Neural Operator}
\label{sec:no overview}
Consider the generic form of PDE:
\begin{equation}
    \begin{aligned}
        (\mathcal{L}_a u)(\boldsymbol{x}) &= f(\boldsymbol{x}),  &\boldsymbol{x}\in D \\
        u(\boldsymbol{x}) &= 0, &\boldsymbol{x} \in \partial D,
    \end{aligned}
\end{equation}
where $a \in \mathcal{A}$, $f \in \mathcal{U}^*$ with $d$-dimensional solution domain $D \in \mathbb{R}^d$ and boundary $\partial D$. We assume that the solution $u: D \rightarrow \mathbb{R}^{d_u}$ and the input parameter $a : D \rightarrow \mathbb{R}^{d_a}$ live in Banach spaces  $\mathcal{U}$ and $\mathcal{A}$, respectively. Expression $\mathcal{L}_a : \mathcal{A} \rightarrow \mathcal{L}(\mathcal{U};\mathcal{U}^*)$ is a mapping from the parameter Banach space $\mathcal{A}$ to the space of linear operators $\mathcal{U}$ to its dual $\mathcal{U}^*$. An operator $\mathcal{G}^{\dagger} := \mathcal{L}_a^{-1}f : \mathcal{A} \rightarrow \mathcal{U}$ is defined to map the input parameter to the solution $a \mapsto u$. Suppose we have a set of $N$ observations $\{ a^{(i)}, u^{(i)} \}^N_{i=1}$, the goal is to build an approximation of $\mathcal{G}^{\dagger}$ by constructing:
\begin{equation}
    \mathcal{G}_{\boldsymbol{\theta}} : \mathcal{A} \rightarrow \mathcal{U}, \quad \boldsymbol{\theta} \in \mathbb{R}^p,
\end{equation}
then optimising paratemer $\boldsymbol{\theta} \in \mathbb{R}^p$, where $p$ is the dimensionality of the parameter, so that $\mathcal{G}_{\boldsymbol{\theta}^\dagger} \approx \mathcal{G}^\dagger$.

Neural operators (NO) \cite{Kovachki2021} are a class of neural network models that are designed to learn grid-independent problems in infinite-dimensional function space, in contrast to the classical neural network models that learn the mapping between finite-dimensional vector spaces. The neural operator model is generally composed of linear integral operators and non-linear activation functions that are developed specifically to tackle the learning problem between Banach spaces. The NO is formulated as an iterative architecture which resembles a multi-layered neural network:
\begin{equation} \label{eq: no_architecture}
    a(\boldsymbol{x}) \mapsto v_0 \mapsto v_1 \mapsto \cdots \mapsto v_{l} \mapsto u(\boldsymbol{x}),
\end{equation}
where $v_j$ for $j=0,1,\ldots,l$ is a sequence of functions that transforms the input to its latent representation. The input $a \in \mathcal{A}$ is initially transformed into a higher dimensional representation $v_0(\boldsymbol{x}) = P(a(\boldsymbol{x}))$ through a local operator $P$ (also known as the lifting layer \cite{Kovachki2021}), which usually a shallow fully-connected NN is used in this task. The output $u(\boldsymbol{x}) = Q(v_l(\boldsymbol{x}))$ corresponds to the projection of the final latent space into the output space, where this step is also typically parameterised by a shallow fully connected NN.

The ability of neural operator models to learn the mappings between infinite-dimensional function spaces is made possible by utilising integral operators inside of each iteration update $v_j \mapsto v_{j+1}$. The update is defined as the non-local integral operator $\mathcal{K}$ and a nonlinear activation function $\sigma$, 
\begin{equation} \label{eq: FNO_layer}
    v_{j+1}(x) := \sigma(\boldsymbol{W}v_j(\boldsymbol{x}) + (\mathcal{K}_j(v_j))(\boldsymbol{x}))
\end{equation}
The linear integral operator is given by.
\begin{equation}
    (\mathcal{K}_j(v_j))(\boldsymbol{x}) = \int_{D_j} \kappa^{(j)}(\boldsymbol{x},\boldsymbol{y})v_j(\boldsymbol{y}) d\nu_j(\boldsymbol{y}) \approx \sum_i^N \kappa(\boldsymbol{x},\boldsymbol{y}_i)v_j(\boldsymbol{y}_i) \Delta y_i,
\end{equation}
where $v_j(\cdot)$ is the input function to the operator block, and $\kappa(x,y)$ is the learnable kernel between two points in the output and input domain. Then, the linear integral operator is followed by non-linear activation functions such as Gaussian error linear units (GeLU) \cite{hendrycks2023gelu}, sigmoid linear units (SiLU)\cite{elfwing2017silu}, or rectified linear units (ReLU) \cite{agarap2019relu}.

The kernel integrator in neural operator models is available in many forms of different design options, such as the graph neural operator (GNO) \cite{li2020gno}, low-rank neural operator (LNO) \cite{Kovachki2021}, Fourier neural operator (FNO) \cite{li2021fno}, and wavelet neural operator (WNO) \cite{Tripura2024}, each offering distinct advantages, disadvantages, and use cases. For the sake of brevity, this paper focuses exclusively on the FNO formulation. However, in some of the experiments, WNO was also implemented.

FNO \cite{li2021fno} uses the Fourier transform as the kernel integral transformation. This method is effective in tackling challenging PDE problems, such as learning solutions in fluid dynamics problems \cite{li2021fno}. To define the kernel integrator in FNO, let $\mathcal{F}$ be the Fourier transform of a function $f:D \rightarrow \mathbb{R}^{d_v}$ and $\mathcal{F}^{-1}$ is its inverse, 
\begin{equation}
    \begin{aligned}
        (\mathcal{F}f)_j(\boldsymbol{k}) &= \int_D f_j(\boldsymbol{x})e^{-2i\pi (\boldsymbol{x},\boldsymbol{k})} dx \\
        (\mathcal{F}^{-1}f)_j(\boldsymbol{x}) &= \int_D f_j(\boldsymbol{k})e^{2i\pi (\boldsymbol{x},\boldsymbol{k})} d\boldsymbol{k},
    \end{aligned}
\end{equation}
for $j = 1, \ldots, d_v$ where $i=\sqrt{-1}$ is the imaginary unit. Thus, we define the Fourier kernel integrator as:
\begin{equation}
    (\mathcal{K}(a;\phi)v_t)(\boldsymbol{x}) = \mathcal{F}^{-1}(\mathcal{F}(\kappa_\phi)\cdot \mathcal{F}(v_t))(\boldsymbol{x}), \  \forall \boldsymbol{x} \in D,
\end{equation}
where $\kappa_\phi$ is parameterised in Fourier space.

When defining the model, we typically retain only the first $N_m$ modes, where $N_m$ is a user-defined hyperparameter. This means that we are not using all Fourier modes from the Fourier transform result; instead, we discard the high-frequency modes. Choosing the right number of $N_m$ might be a unique problem for FNO. Since the higher frequency modes are interpreted as the non-linearity term in the model, one might suggest that we keep high enough amounts of modes in the model. Thus, this results in more efficient computation without sacrificing too much accuracy.

\begin{figure}[H]
    \centering
    \includegraphics[width=0.75\textwidth]{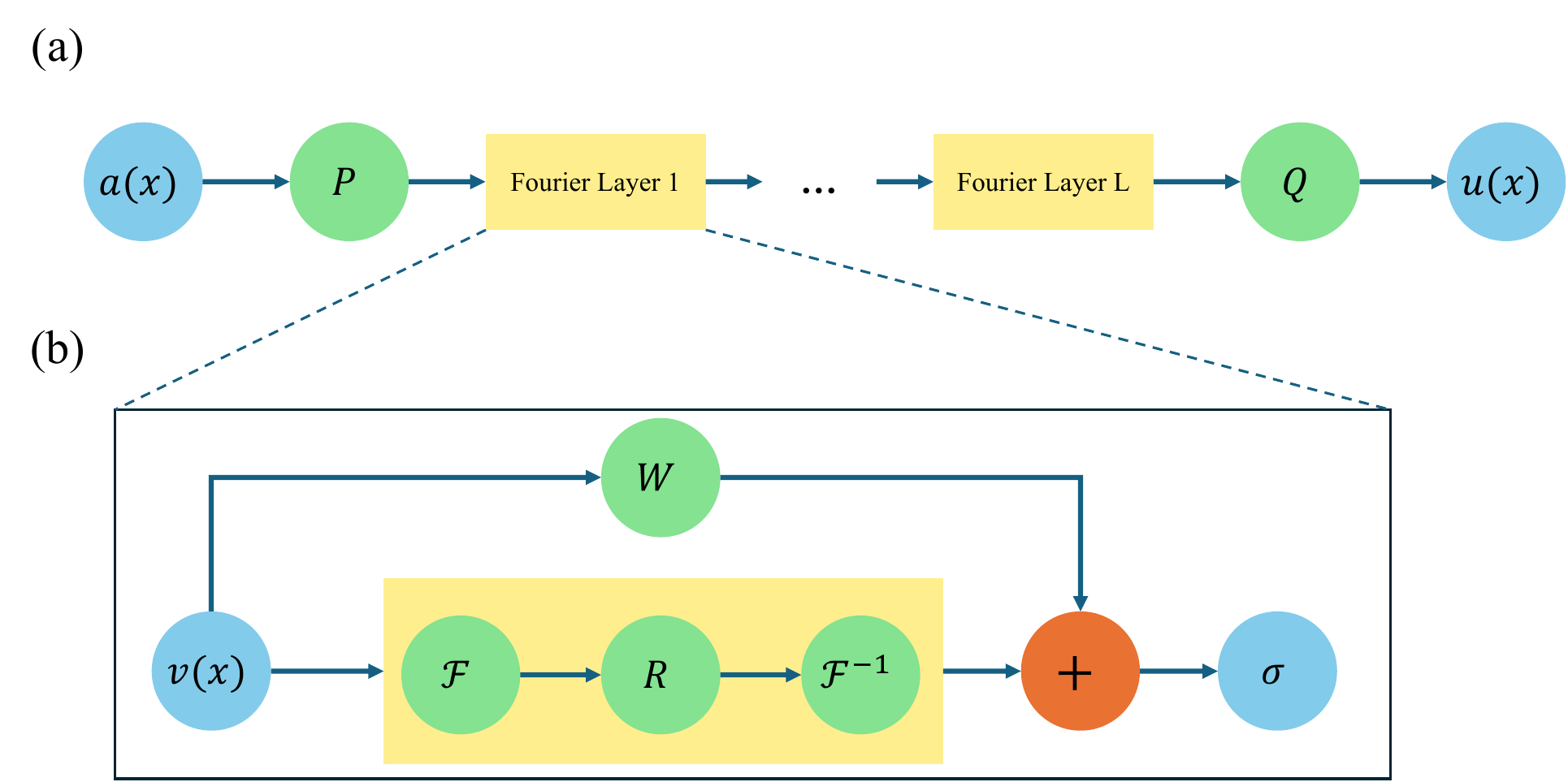}
    \caption{FNO schematics: (a) full architecture of FNO. (b) Fourier layers.}
    \label{fig: FNO schematics}
\end{figure}

The schematic figure of the FNO is given in Fig. \ref{fig: FNO schematics}. The full architecture of the neural operator (a) resembles what we have described in Eq. (\ref{eq: no_architecture}), and the Fourier integral kernel from Eq. (\ref{eq: FNO_layer}) is illustrated in Fig. \ref{fig: FNO schematics}(b). The bottom part of the schematic \ref{fig: FNO schematics}(b) describes the Fourier integral kernel, and the top part illustrates the bias. The operation $R$ in the Fourier integral kernel denotes that we retain the first $N_m$ modes from the Fourier transformation operation. It is also worth noticing that the bias $W$ in Fig. \ref{fig: FNO schematics}(b) is modelled as $d$-dimensional $1 \times 1$ convolutional layer, where $d$ refers to the spatial dimension of the problem.

\subsection{Multi-fidelity learning}
\label{sec:mf overview}
Multiple levels of simulation fidelity are often found in PDE-based systems. The goal of multi-fidelity simulation lies in balancing accuracy and computational efficiency. High-fidelity (HF) simulations provide detailed and highly accurate results but require significant computational resources. In contrast, low-fidelity (LF) simulations are more computationally efficient, though they trade off some degree of accuracy and detail. In engineering applications, the availability of multi-level simulation fidelity has been exploited to assist the process of design optimisation \cite{Koziel2014, Liu2022}. In this matter, several methods have been implemented to perform multi-fidelity modelling, such as model fusion \cite{Eldred2004}, space mapping \cite{Robinson2008}, and shape-preserving response prediction \cite{Leifsson2010}.

Before the utilisation of neural networks as a surrogate model became popular due to the advent of computational resources, the multi-fidelity learning approach has been implemented in other popular surrogate models such as linear regression \cite{Zhang2018, FernndezGodino2019}, Gaussian process regression \cite{Myers1982, Forrester2007}, and polynomial chaos expansion (PCE) regression \cite{Palar2016}. Despite different multi-fidelity surrogate model (MFSM) methods, approaches, and applications that have been developed, the core concept of multi-fidelity learning in surrogate models is the exploitation of the correlation between the HF and LF data. One of the most used representations is the comprehensive correction \cite{GiselleFernndezGodino2023}, described as:
\begin{equation}
    \hat{\boldsymbol{y}}_{HF} = \rho(\boldsymbol{x}) \cdot \boldsymbol{y}_{LF}(x) + \delta(\boldsymbol{x}),
\end{equation}
where $\rho(\boldsymbol{x})$ represents the multiplicative correlation between multi-fidelity data, and $\delta(\boldsymbol{x})$ represents the additive correlation. The problem with this representation is that it only handles linear correlations between the lower and higher fidelity. However, the correlation between LF and HF data might go beyond linear correlation, such as the mixed convection flows past a cylinder \cite{Babaee2016, Perdikaris2017}. To capture the nonlinear correlation, a generalised autoregressive scheme is proposed in \cite{Meng2020}, which is expressed as:
\begin{equation}\label{eq: nl-mapping}
    \hat{\boldsymbol{y}}_{HF} = F(\boldsymbol{y}_{LF}) + \delta(\boldsymbol{x}),
\end{equation}
where $F(\cdot)$ is an unknown, possibly nonlinear, function that maps between the low and the high-fidelity data. In terms of function mapping, Eq. (\ref{eq: nl-mapping}) rewrites as:
\begin{equation}
    \hat{\boldsymbol{y}}_{HF} = \mathcal{F}(\boldsymbol{x},\boldsymbol{y}_{LF};\theta).
\end{equation}
In this case, \cite{Meng2020} use a neural network to represent $\mathcal{F}(\cdot)$.

To represent the non-linear mapping function with a neural network, the mapping function $\mathcal{F}(\cdot)$ is then decomposed into linear and nonlinear parts:
\begin{equation}\label{eq: decompose_f}
    \mathcal{F} = \mathcal{F}_{l} + \mathcal{F}_{nl},
\end{equation}
where $\mathcal{F}_l$ and $\mathcal{F}_{nl}$ are the linear and nonlinear terms in $\mathcal{F}$, respectively. Thus, the correlation is now written as:
\begin{equation} \label{eq: decompose_f2}
    \hat{\boldsymbol{x}}_{HF} = \gamma \mathcal{F}_{l}(\boldsymbol{x},\boldsymbol{x}_{LF};\boldsymbol{\theta}) + (1-\gamma) \mathcal{F}_{nl}(\boldsymbol{x},\boldsymbol{y}_{LF};\boldsymbol{\theta}), \ \gamma \in [0,1],
\end{equation}
where $\gamma$ is a user-defined hyperparameter.

Similarly, Guo et al. \cite{Guo2022} proposed a multi-fidelity neural network architecture. In the proposed architecture, instead of decomposing the mapping function $\mathcal{F}(\cdot)$ to the linear and nonlinear terms as shown in Equation (\ref{eq: decompose_f}), they directly model the function $\mathcal{F}(\boldsymbol{x},\boldsymbol{y}_{LF};\boldsymbol{\theta})$. This architecture uses the same input layer for the LF and HF data. Then, at the end of the network, the output layer of the HF is given, and the output layer of the LF data is located in the middle of the hidden layers. Thus, this architecture is called the "intermediate architecture". Since the LF and HF data are processed simultaneously in a single pass, two different training loss components are introduced, namely the ${MSE}_{LF}$ and ${MSE}_{HF}$:
\begin{equation}\label{eq: loss_intermediate1}
    \begin{aligned}
        & MSE_{LF} = \frac{1}{N_{LF}} \sum_{i=1}^{N_{LF}} \left( \boldsymbol{y}_{LF}^{(i)} - \hat{\boldsymbol{y}}_{LF}(\boldsymbol{x}_{LF}^{(i)})\right)^2, \ \text{and} \\  
        & MSE_{HF} = \frac{1}{N_{HF}} \sum_{i=1}^{N_{HF}} \left( \boldsymbol{y}_{HF}^{(i)} - \hat{\boldsymbol{y}}_{HF}(\boldsymbol{x}_{HF}^{(i)})\right)^2.
    \end{aligned} 
\end{equation}
Where $\boldsymbol{y}^{(i)}$ indicates the ground truth and $\hat{\boldsymbol{y}}(\boldsymbol{x}^{(i)})$ indicates the predicted output $\hat{\boldsymbol{y}}$ given $\boldsymbol{x}$. The subscripts $LF$ and $HF$ are used to indicate the low fidelity and high fidelity data, respectively. A weighted sum is then used to aggregate the two loss functions to account for objectives of different scales. Thus, the loss function in this architecture is written as:
\begin{equation}\label{eq: loss_intermediate2}
    \mathcal{L} = \alpha MSE_{HF} + (1-\alpha) MSE_{LF} + \lambda ||\boldsymbol{\theta}||^2,
\end{equation}
where $\alpha \in [0,1]$ is the weighting factor of the loss function and $\lambda ||\boldsymbol{\theta}||^2$ acts as the regularisation term of the parameters $\boldsymbol{\theta}$.

\subsubsection{Intermediate architecture}
\label{sec: Intermediate architecture}
\begin{figure}[H]
    \centering
    \includegraphics[width=0.7\textwidth]{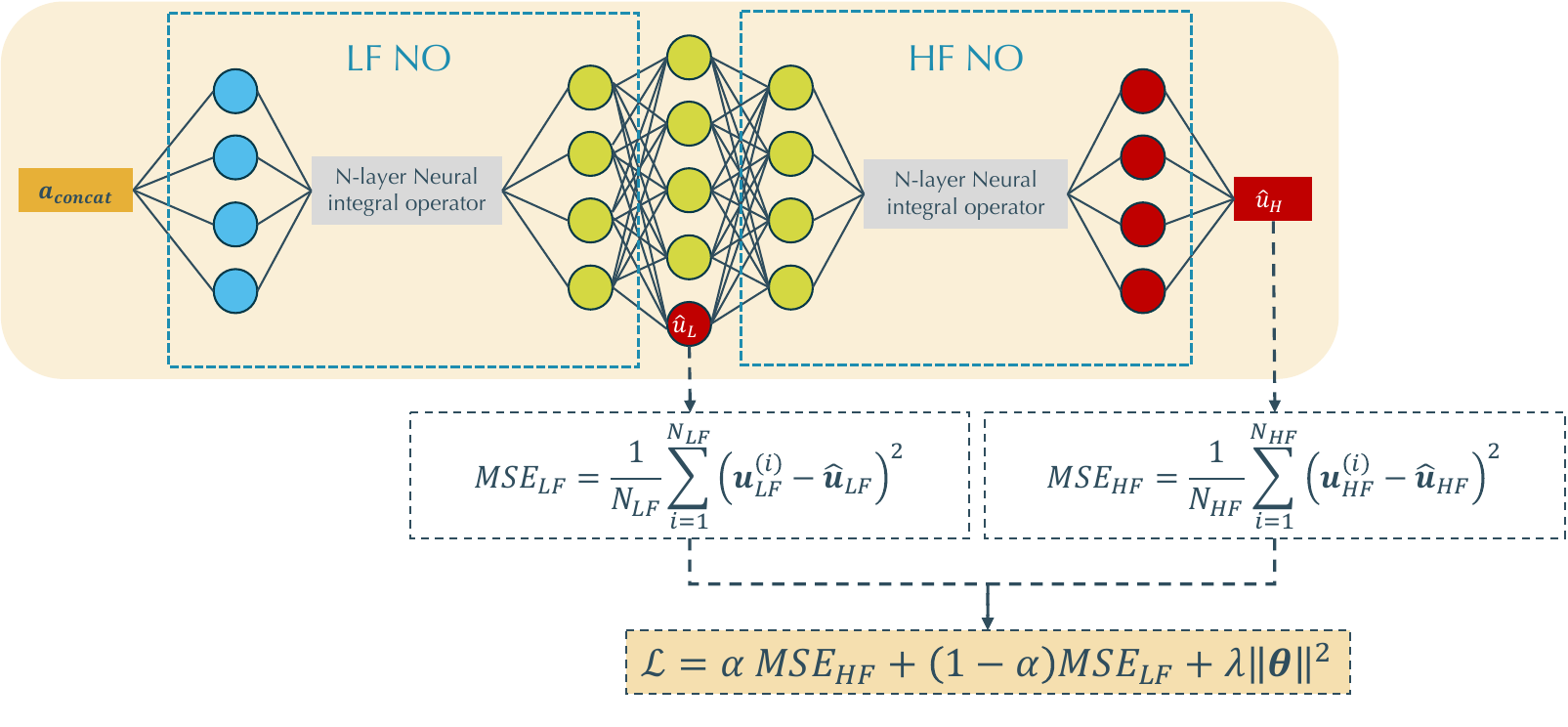}
    \caption{Intermediate multi-fidelity neural operator architecture}
    \label{fig: Intermediate FNO}
\end{figure}

Inspired by \cite{Guo2022}, we implement the intermediate multi-fidelity neural network architecture in the neural operator. Illustrated in Fig. \ref{fig: Intermediate FNO}, the intermediate multi-fidelity neural operator (MFNO) consists of 2 neural operators that are connected with a fully connected layer in which one of the nodes in the layer is the lower-fidelity output. The model is trained by minimising the loss function defined in Eq. (\ref{eq: loss_intermediate2}), where the $\alpha$ is a user-defined hyperparameter since the lower and higher-fidelity neural operator shares the same input layer. Both training datasets $\mathcal{T}_{LF} = \{ (\boldsymbol{x}_{LF}^{(i)}, \boldsymbol{y}_{LF}^{(i)}): 1 \leq i \leq N_{LF} \}$ and $\mathcal{T}_{HF} = \{ (\boldsymbol{x}_{HF}^{(i)}, \boldsymbol{y}_{HF}^{(i)}): 1 \leq i \leq N_{HF} \}$ are concatenated before training the model.

\subsubsection{Multi-step architecture}
\begin{figure}[H]
    \centering
    \includegraphics[width=0.65\textwidth]{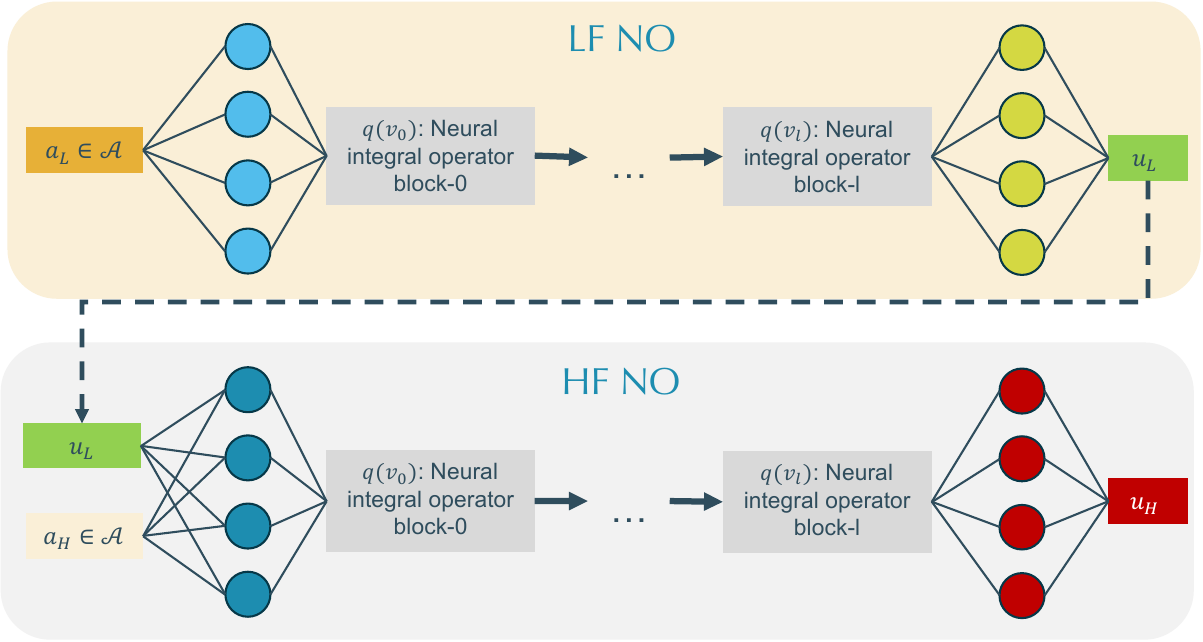}
    \caption{Two-step multi-fidelity neural operator architecture}
    \label{fig: 2steps FNO}
\end{figure}

 Another approach, also used to define a multi-fidelity neural operator is by employing two distinct neural operators to model these functions. In this multi-step \cite{Guo2022} approach, the first neural operator $NO_{LF}$ is trained on the lower-fidelity training set $\mathcal{T}_{LF}$. Then a second neural operator $NO_{HF}$ approximates the high-fidelity function $\hat{\boldsymbol{y}}_{HF} = \mathcal{F}_{HF}(\boldsymbol{x}_{HF},\boldsymbol{y}_{LF})$ based on the HF input $\boldsymbol{x}_{HF}$ and the LF output $\hat{\boldsymbol{y}}_{LF}$ at the same locations with $\boldsymbol{y}_{HF}$. Since the LF and HF data may be generated independently, the $\boldsymbol{y}_{LF}$ at location $\boldsymbol{x}_{HF}$ might not be directly available. Thus, the data has to be evaluated from the lower fidelity neural operator, $\hat{\boldsymbol{y}}_{LF} = \mathcal{F}_{LF}(\boldsymbol{x}_{HF})$. The implementation of the two-step multi-fidelity neural operator is illustrated in Fig. \ref{fig: 2steps FNO}.

\begin{figure}[H]
    \centering
    \includegraphics[width=0.75\textwidth]{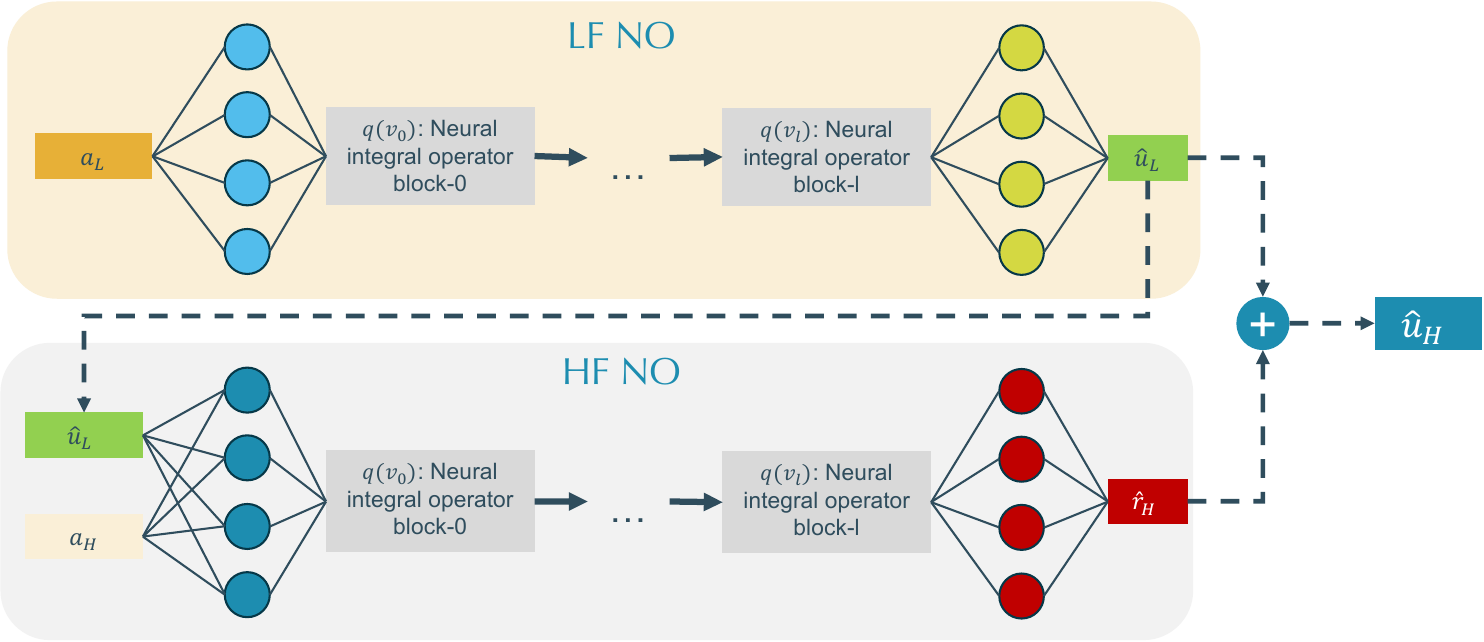}
    \caption{Residual multi-fidelity neural operator architecture}
    \label{fig: residual NO}
\end{figure}

An alternative form of two-step multi-fidelity learning is based on the residual learning approach. Proposed in \cite{Tripura2024}, the residual MFNO use a similar architecture to the two-step MFNO as shown in Fig. \ref{fig: 2steps FNO}. However, the second neural operator learns the residual value instead of the high-fidelity output, defined as:
\begin{equation}
    r(\boldsymbol{x}) = \boldsymbol{y}_{HF}(\boldsymbol{x}) - \boldsymbol{y}_{LF}(\boldsymbol{x}).
\end{equation}
The goal of using residual learning, rather than directly mapping to the high-fidelity (HF) solution, is to simplify the learning task. Since there is generally some correlation between the low-fidelity (LF) and HF solutions, a degree of feature similarity can be expected. In some cases, however, this correlation may be weak or dominated by a systematic bias. In such scenarios, learning the residual, the difference between the HF and LF outputs, allows the model to focus on capturing the remaining complexity not already represented by the LF solution. This approach can lead to improved performance and data efficiency, as supported by \cite{Tripura2024}. The residual MFNO architecture is depicted in Fig. \ref{fig: residual NO}.

\subsubsection{Transfer learning}
\begin{figure}[H]
    \centering
    \includegraphics[width=0.6\textwidth]{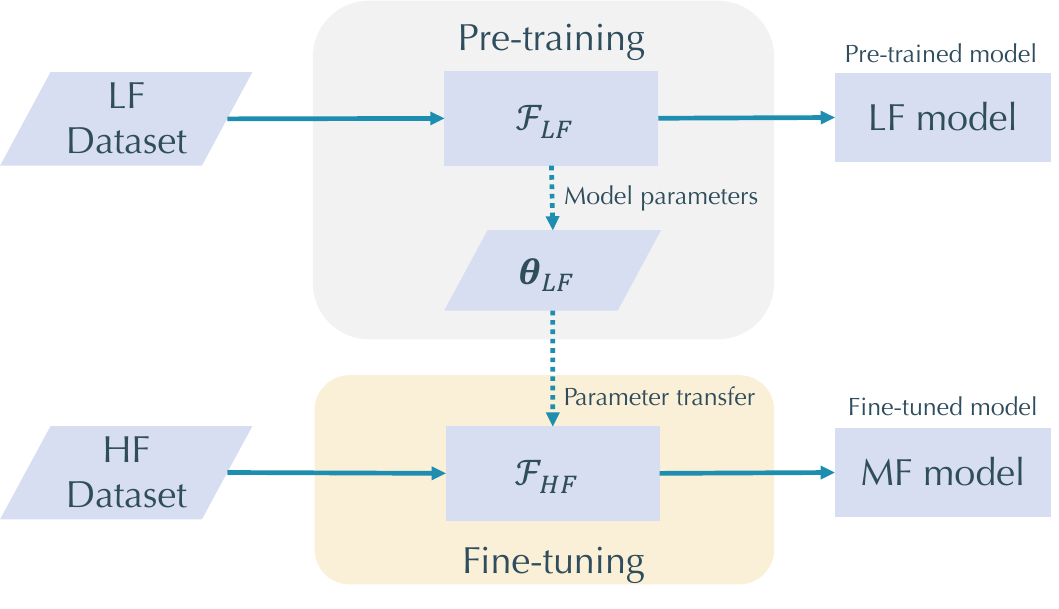}
    \caption{Transfer learning process schematics.}
    \label{fig: TF NO}
\end{figure}

Another widely used approach in multi-fidelity modelling is transfer learning \cite{Lyu_2023, tang2024}. The central idea is to leverage knowledge gained from training on low-fidelity (LF) data to improve performance on high-fidelity (HF) tasks. This is achieved through a two-stage process: pre-training and fine-tuning. During the pre-training phase, a model is trained using a large set of inexpensive LF data $\mathcal{T}_{LF}$ following a standard neural operator training procedure \cite{Kovachki2021}  yielding a set of optimum LF NO parameters $\boldsymbol{\theta}_{LF}$. This set of optimum LF NO parameters is then transferred to the initialised HF NO as a starting point for the fine-tuning process. A limited amount of expensive, high-fidelity data is then employed to train the HF NO in the fine-tuning process. After this fine-tuning process, we then obtain the multi-fidelity model in the form of the fine-tuned HF NO model. The schematic of the transfer learning process is given in Fig. \ref{fig: TF NO}.

\section{Experiment results}
\label{sec:experiments}
We present four test cases. The first two, taken from \cite{Tripura2024, Lu2022}, are the 1-dimensional stochastic Poisson equation and 2-dimensional Darcy flow in a triangular domain act as a benchmark. Then, we present the modified version of the 2-dimensional Darcy flow as our third test case. Finally, we will evaluate the models based on the unsteady smoke inflow problem \cite{ZakariaSmoke2024}. The main goal of this test case is to assess the multi-fidelity model performance on time-series data with a complex problem, in this case, an unsteady laminar flow.

\subsection{1-dimensional stochastic Poisson equation}
\label{sec: 1dpoisson}
\begin{figure}[H]
    \centering
    \includegraphics[width=0.7\textwidth]{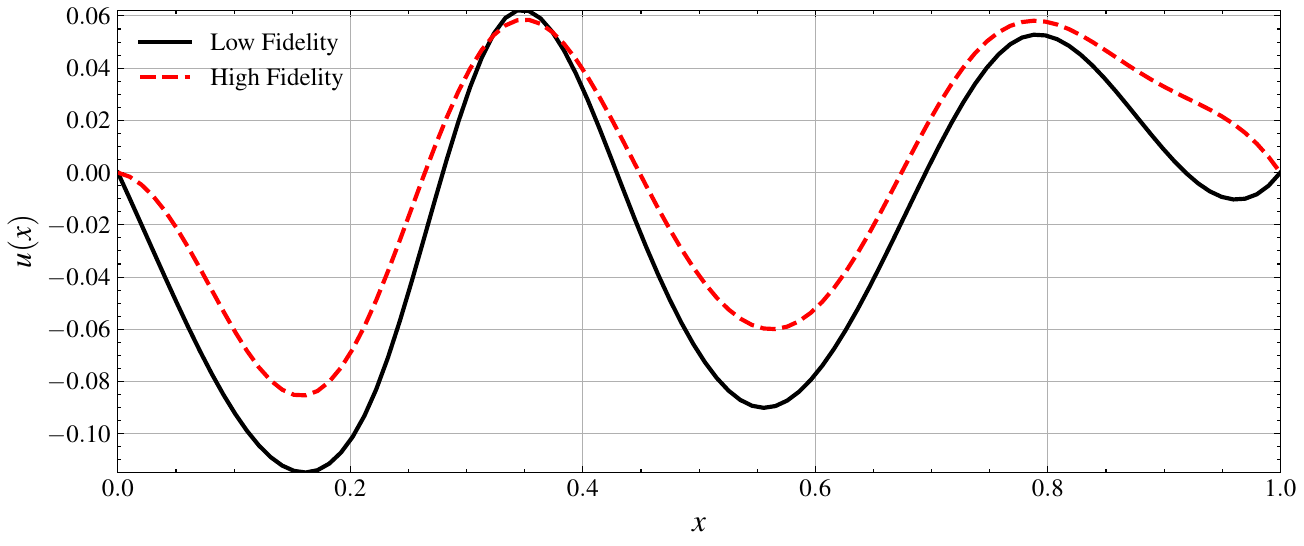}
    \caption{An example of the difference between LF and HF data in a 1-dimensional stochastic Poisson equation.}
    \label{fig: 1D_lfvshf}
\end{figure}

Following \cite{Tripura2024}, we consider the second-order stochastic Poisson's equation in one dimension as our first test case. The stochastic differential equation (SDE) is formulated as:
\begin{equation}
    \begin{aligned}
        && \frac{d^2 u(x)}{dx^2} = 20 g(x), \  x\in [0,1], \\
        && u(x=0) = u(x=1) = 0, \\
        && g(x) \sim \mathcal{GP}(m(x), k(x,x')),
    \end{aligned}
    \label{eq: 1D SPDE}
\end{equation}
where $g(x)$ is the spatially random forcing function, modelled as a Gaussian process ($\mathcal{GP}$) with mean $m(x)$ set to be 0, and the radial basis function kernel $k(x,x') = \exp \left( -(x-x')^2 / (2l^2) \right)$ where the kernel length scale of the GP is set to be $l=0.1$. The HF dataset is generated using a grid size $\Delta x =0.01$, while the LF dataset grid size is  $\Delta x =0.1$. An illustration of the differences between LF and HF data is presented in Fig.~\ref{fig: 1D_lfvshf}. As shown, the LF data captures the overall trend of the HF solution but lacks finer details present in the more accurate HF output. 

We vary the number of training samples on the LF and HF models to investigate the effect of the number of training samples on the multi-fidelity model. Each experiment is repeated 15 times with different combinations of testing and training dataset splits. We found that the FNO yields a better performance compared to other kernels, such as the Wavelet neural operator (WNO), in this test case. However, since the selection of the NO kernel is not the focus of this paper, the comparison between kernels is not shown in this paper. The LF and HF model hyperparameters are given in Table \ref{tab:1D FNO hyperparams}. The choice of the hyperparameters is arbitrary. In this experiment, we found that the hyperparameters in Table \ref{tab:1D FNO hyperparams} yield a balance between accuracy and computational cost.

\begin{table}[!ht]
    \centering
    \caption{1-dimensional multi-fidelity FNO hyperparameters.}
    \begin{tabular}{l l l}
        \hline
         & Lower-fidelity & Higher-fidelity \\
        \hline
        NO kernel & Fourier & Fourier \\
        Number of NO layers & 4 & 2 \\
        FC layers width & 64 & 64 \\
        Fourier modes & 16 & 16 \\
        Learning rate& $1\times 10^{-3}$ & $1\times 10^{-3}$ \\
        Weight Decay & $1\times 10^{-4}$ & $1\times 10^{-4}$ \\
        Epochs & 200 & 200 \\
        \hline
    \end{tabular}
    \label{tab:1D FNO hyperparams}
\end{table}

The experiments were conducted mainly to investigate the behaviour of the models given different training samples in both LF and HF. The result in Table \ref{tab:1D FNO results} is interpreted as two experiments. First, we set a constant value for the number of HF training samples to 50 and gradually increased the LF training sample number from 50 to 150. Second, we set $n_{\text{train}}$ LF to be constant at 150 while gradually increasing HF from 15 to 50. While we have access to many HF data, we keep the highest number of HF training samples to 50 to emulate real-world multi-fidelity cases. In this test case, we provide the high-fidelity baseline, which is trained only on the corresponding high-fidelity test sample.

\begin{figure}[H]
    \centering
    \includegraphics[width=0.65\textwidth]{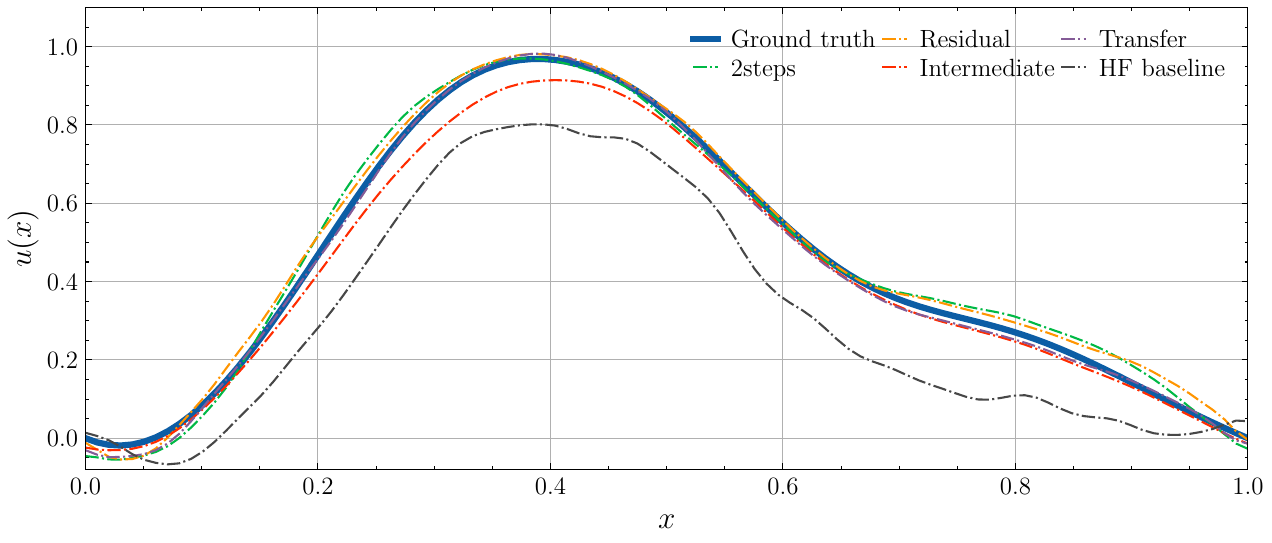}
    \caption{Single realisation on 1-dimensional stochastic Poisson equation test data.}
    \label{fig: 1D realization}
\end{figure}

The result shows that the intermediate architecture is inferior compared to the 2-step and residual architectures. The 2-step and residual architecture seem to have their advantages in particular cases by a small margin from each other. Meanwhile, the transfer learning, while not giving an outstanding performance in this test case, has the most robust results concerning the LF and HF variation. A single realisation of the prediction on the test dataset is given in Fig. \ref{fig: 1D realization}

\begin{table}[!ht]
\caption{1-dimensional multi-fidelity FNO RMSE}
\resizebox{\textwidth}{!}{%
    \begin{tabular}{lllllll}
        \hline
        \multicolumn{2}{l}{$n_{\text{train}}$} & \multirow{2}{*}{2-steps} & \multirow{2}{*}{Residual} & \multirow{2}{*}{Intermediate} & \multirow{2}{*}{Transfer learning} & \multirow{2}{*}{HF baseline}\\ \cline{1-2}
        LF  & HF & & & & \\ \hline
        50  & 50 & $\mathbf{\num{4.43e-2} \pm \num{9.35e-3}}$ & $\num{6.83e-2} \pm \num{1.49e-2}$ & $\num{5.60e-2} \pm \num{1.21e-2}$ & $\num{4.80e-2} \pm \num{1.06e-2}$ & $\num{5.71e-2} \pm \num{1.23e-2}$ \\
        100 & 50 & $\mathbf{\num{3.07e-2} \pm \num{5.96e-3}}$ & $\num{3.54e-2} \pm \num{7.40e-3}$ & $\num{5.43e-2} \pm \num{1.42e-2}$ & $\num{3.85e-2} \pm \num{9.68e-3}$ & $\num{5.71e-2} \pm \num{1.23e-2}$ \\
        150 & 50 & $\mathbf{\num{2.58e-2} \pm \num{3.55e-3}}$ & $\num{2.69e-2} \pm \num{7.06e-3}$ & $\num{7.97e-2} \pm \num{1.66e-2}$ & $\num{3.40e-2} \pm \num{6.88e-3}$ & $\num{5.71e-2} \pm \num{1.23e-2}$ \\
        150 & 25 & $\num{3.93e-2} \pm \num{1.00e-2}$ & $\mathbf{\num{3.31e-2} \pm \num{6.33e-3}}$ & $\num{6.84e-2} \pm \num{2.13e-2}$ & $\num{3.85e-2} \pm \num{7.92e-3}$ & $\num{1.24e-1} \pm \num{3.02e-2}$ \\
        150 & 15 & $\num{5.99e-2} \pm \num{1.07e-2}$ & $\mathbf{\num{3.95e-2} \pm \num{4.91e-3}}$ & $\num{7.83e-2} \pm \num{2.92e-2}$ & $\num{3.96e-2} \pm \num{6.01e-3}$ & $\num{2.11e-1} \pm \num{5.67e-2}$ \\ \hline
    \end{tabular}}
    \label{tab:1D FNO results}
\end{table}

\subsection{2-dimensional triangular Darcy flow}
\label{sec: 2ddarcy}
The Darcy equation is often used to describe flow through a porous medium in geotechnical and civil engineering. Following \cite{Tripura2024, Lu2022}, we consider this problem in the 2-dimensional triangular domain with a notch. The mathematical expression of the Darcy equation is given by:
\begin{equation}
    \begin{gathered}
    -\nabla \cdot \left(a(x,y)\nabla u(x,y) \right) = f(x,y); \ x,y \in [0,1],\\
    u(x,y)|_{\partial D} = \mathcal{GP} \left(0, \mathcal{K}\left((x,y),(x',y') \right) \right)
    \end{gathered}
    \label{eq: Darcy}
\end{equation}
where $a(x,y) \in \mathbb{R}$ is the permeability field, $u(x,y) \in \mathbb{R}$ is the pressure field, and $f(x,y) \in \mathbb{R}$ is the source term. The synthetic training and testing data are generated by solving Eq. (\ref{eq: Darcy}) for different boundary conditions. Different boundary conditions $u(x,y)|_{\partial D}$ are generated from a Gaussian process with radial basis function kernel, defined as:
\begin{equation}
    \mathcal{K}\left((x,y),(x',y') \right) = \exp \left( \frac{-(x-x')^2}{2l^2_x} - \frac{(y-y')^2}{2l^2_y} \right),
\end{equation}
where the lengthscale parameter $l_x$ and $l_y$ are set as 0.2. The permeability and the source field are assumed constant with $a(x,y)=0.1$ and $f(x,y)=-1$ respectively. 

\begin{figure}[H]
    \centering
    \includegraphics[width=0.6\textwidth]{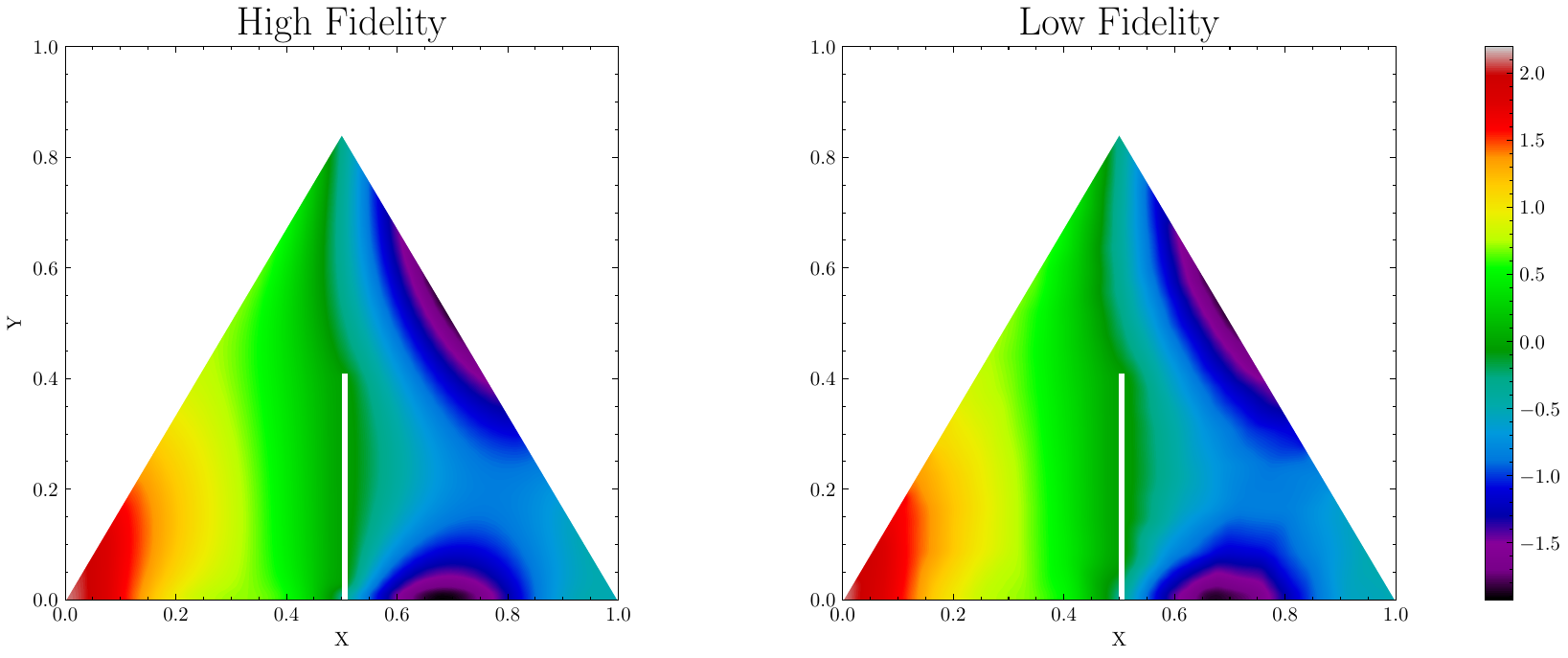}
    \caption{An example of the difference between LF and HF data in the original 2-dimensional Darcy flow problem.}
    \label{fig: 2D_lfvshf}
\end{figure}

Lu et al. \cite{Lu2022} used the MATLAB PDE Toolbox to solve the problem numerically. However, in this paper, we use the generated dataset by Tripura et al. \cite{Tripura2024}, where they already modified the original problem to be suited for a multi-fidelity problem. The different levels of fidelity are achieved through variation in the discretisation grid, where the HF dataset is obtained from a fine grid with maximum and minimum element edge lengths set to $h_{max} = 0.028$ and $h_{min} = 0.026$. Meanwhile, the LF dataset is obtained from a coarser grid with maximum and minimum element edge lengths set to $h_{max} = 0.18$ and $h_{min} = 0.16$. However, as shown in Fig.~\ref{fig: 2D_lfvshf}, the difference between LF and HF results is not significant. 

To train the model, we use the model hyperparameters as given in Table \ref{tab:2D WNO hyperparams}. Here, we follow the WNO hyperparameters used in \cite{Tripura2024}, where the biorthogonal filter is set to \textit{near\_sym\_a} and the quarter shift filter is set to \textit{qshift\_a}. However, the other hyperparameter settings are arbitrary. Similar to section \ref{sec: 1dpoisson}, we conduct the experiments across different combinations of LF and HF training samples. We repeat the experiments with 10 different combinations of testing and training dataset splits for consistency.

\begin{table}[!ht]
    \centering
    \caption{2-dimensional multi-fidelity WNO hyperparameters.}
    \begin{tabular}{l l l}
        \hline
         & Lower-fidelity & Higher-fidelity \\
        \hline
        NO kernel & Wavelet & Wavelet \\
        Number of NO layers & 4 & 4 \\
        FC layers width & 64 & 64 \\
        Decomposition level & 2 & 2 \\
        Biorthogonal filters & near\_sym\_a & near\_sym\_a\\
        Quarter shift filters & qshift\_a & qshift\_a\\
        Learning rate& $1\times 10^{-3}$ & $1\times 10^{-3}$ \\
        Weight Decay & $1\times 10^{-4}$ & $1\times 10^{-4}$ \\
        Epochs & 200 & 200 \\
        \hline
    \end{tabular}
    \label{tab:2D WNO hyperparams}
\end{table}

\begin{figure}[H]
    \centering
    \includegraphics[width=0.85\textwidth]{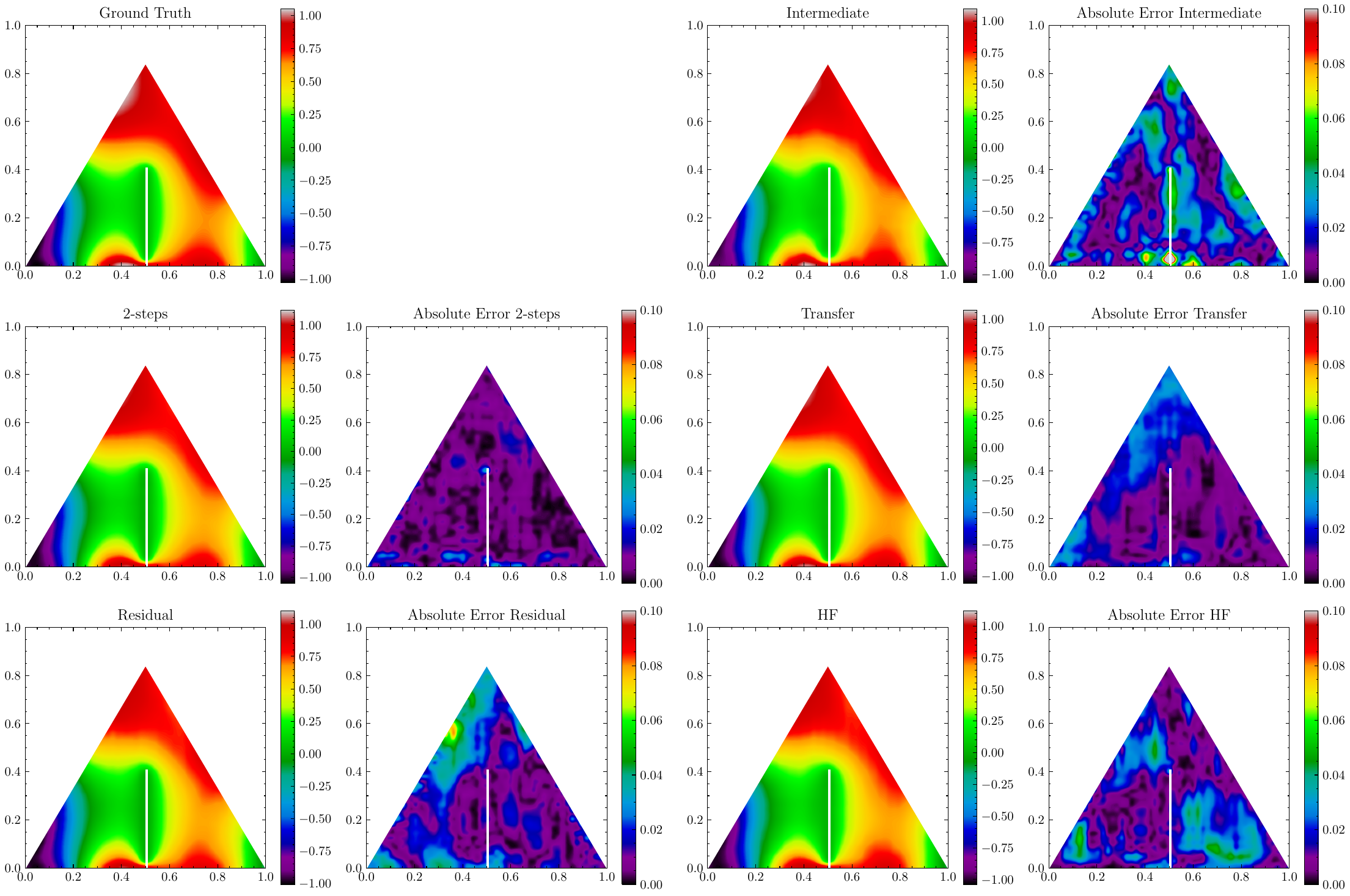}
    \caption{Prediction and error plot of a single data point for each model on 2D Darcy flow in a triangular domain with a notch.}
    \label{fig: Darcy_realization}
\end{figure}

The results in Table \ref{tab:2D WNO results} indicate that transfer learning and the two-step architecture outperform all other methods and are the only approaches that consistently surpass the high-fidelity (HF) baseline across all training sample configurations. We observe that  when high-fidelity data is limited, leveraging lower-fidelity data significantly enhances prediction accuracy. However, when sufficiently high-fidelity training data is available, in this case 50, incorporating lower-fidelity data does not always improve prediction accuracy for residual and intermediate models. This provides us with an insight that while multi-fidelity is promising, we have to be careful with the model and hyperparameter selection. Otherwise, instead of enhancing the model performance, we could end up degrading the prediction model. Fig. \ref{fig: Darcy_realization} presents a realisation plot illustrating these findings. It is important to note that, although some models outperform others, the overall prediction results remain fairly accurate. This is evidenced by the errors, which are all within the same order of magnitude, as shown in Table \ref{tab:2D WNO results}. Furthermore, the error patterns presented in Fig. \ref{fig: Darcy_realization} appear to be purely random, suggesting no systematic bias across the models.

\begin{table}[!ht]
\caption{2-dimensional Darcy flow multi-fidelity WNO RMSE}
\resizebox{\textwidth}{!}{%
    \begin{tabular}{lllllll}
        \hline
        \multicolumn{2}{l}{$n_{\text{train}}$} & \multirow{2}{*}{2-steps} & \multirow{2}{*}{Residual} & \multirow{2}{*}{Intermediate} & \multirow{2}{*}{Transfer learning} & \multirow{2}{*}{HF baseline} \\ \cline{1-2}
        LF  & HF & & & & \\ \hline
        50  & 50 & $\mathbf{\num{2.28e-2} \pm \num{2.35e-3}}$ & $\num{3.50e-2} \pm \num{2.25e-3}$ & $\num{3.64e-2} \pm \num{3.56e-3}$ & $\num{2.32e-2} \pm \num{2.49e-3}$ & $\num{3.14e-2} \pm \num{4.54e-3}$ \\
        100 & 50 & $\num{2.25e-2} \pm \num{3.38e-3}$ & $\num{3.18e-2} \pm \num{2.10e-3}$ & $\num{3.91e-2} \pm \num{2.46e-3}$ & $\mathbf{\num{1.96e-2} \pm \num{1.00e-3}}$ & $\num{3.14e-2} \pm \num{4.54e-3}$ \\
        150 & 50 & $\num{2.23e-2} \pm \num{4.34e-3}$ & $\num{2.63e-2} \pm \num{1.76e-3}$ & $\num{3.67e-2} \pm \num{3.79e-3}$ & $\mathbf{\num{1.74e-2} \pm \num{1.59e-3}}$ & $\num{3.14e-2} \pm \num{4.54e-3}$ \\
        150 & 25 & $\num{2.77e-2} \pm \num{4.55e-3}$ & $\num{2.76e-2} \pm \num{1.95e-3}$ & $\num{5.00e-2} \pm \num{4.50e-3}$ & $\mathbf{\num{2.49e-2} \pm \num{3.13e-3}}$ & $\num{8.86e-2} \pm \num{1.81e-2}$ \\
        150 & 15 & $\num{2.90e-2} \pm \num{8.80e-3}$ & $\num{2.98e-2} \pm \num{8.48e-3}$ & $\num{5.68e-2} \pm \num{1.32e-2}$ & $\mathbf{\num{2.70e-2} \pm \num{3.59e-3}}$ & $\num{2.27e-1} \pm \num{4.32e-2}$ \\ \hline
    \end{tabular}}
    \label{tab:2D WNO results}
\end{table}

\subsection{Modified 2-dimensional triangular Darcy flow}
\label{sec: mod2ddarcy}
The standard Darcy flow problem described in Section \ref{sec: 2ddarcy} assumes that multi-fidelity data is obtained through differences in mesh resolution. In this setup, lower-fidelity data is generated by solving the same numerical formulation on a coarser grid or mesh, resulting in a mean absolute error of 0.02647 between the low-fidelity (LF) and high-fidelity (HF) data, as shown in Fig.~\ref{fig: 2D_lfvshf}. While this approach is a valid form of multi-fidelity learning, real-world multi-fidelity data often exhibits more complex relationships, leading to greater discrepancies between LF and HF solutions. For example, in computational fluid dynamics (CFD) problems, the user can specify different assumptions or turbulence equations, which significantly impact the solution. As an illustration, Figure 6 in \cite{renard2021} highlights the differences between turbulence models such as Reynolds-averaged Navier-Stokes (RANS), delayed detached eddy simulation (DDES), and zonal detached eddy simulation (ZDES), demonstrating how variations in governing equations lead to substantial differences in predicted flow behaviour. 

To emulate the behaviour of multi-fidelity in the governing equation, we propose to modify the 2-dimensional triangular Darcy flow in \cite{Tripura2024}. Instead of decreasing the resolution to achieve lower fidelity, we decompose the solution using a proper orthogonal decomposition (POD) \cite{chatterjee_pod} method. The snapshot matrix $\boldsymbol{A}$ of the solution is decomposed through a lower rank approximation, defined as:
\begin{equation}
    \boldsymbol{A} \approx \boldsymbol{U} \boldsymbol{\Sigma}_k \boldsymbol{V}^T,
\end{equation}
where $\boldsymbol{U}$ is an $N \times k$ orthogonal matrix, $\boldsymbol{V}$ is an $m \times k$ orthogonal matrix, and $\boldsymbol{\Sigma}_k$ is a $k \times k$ matrix with all elements zero except its diagonal.  In this case, we only retain the two most dominant modes, resulting in a reduced-order snapshot matrix $\boldsymbol{A}_2$.
\begin{equation}
    \boldsymbol{A}_2 = \boldsymbol{U} \boldsymbol{\Sigma}_2 \boldsymbol{V}^T
\end{equation}

\begin{figure}[H]
    \centering
    \includegraphics[width=0.75\textwidth]{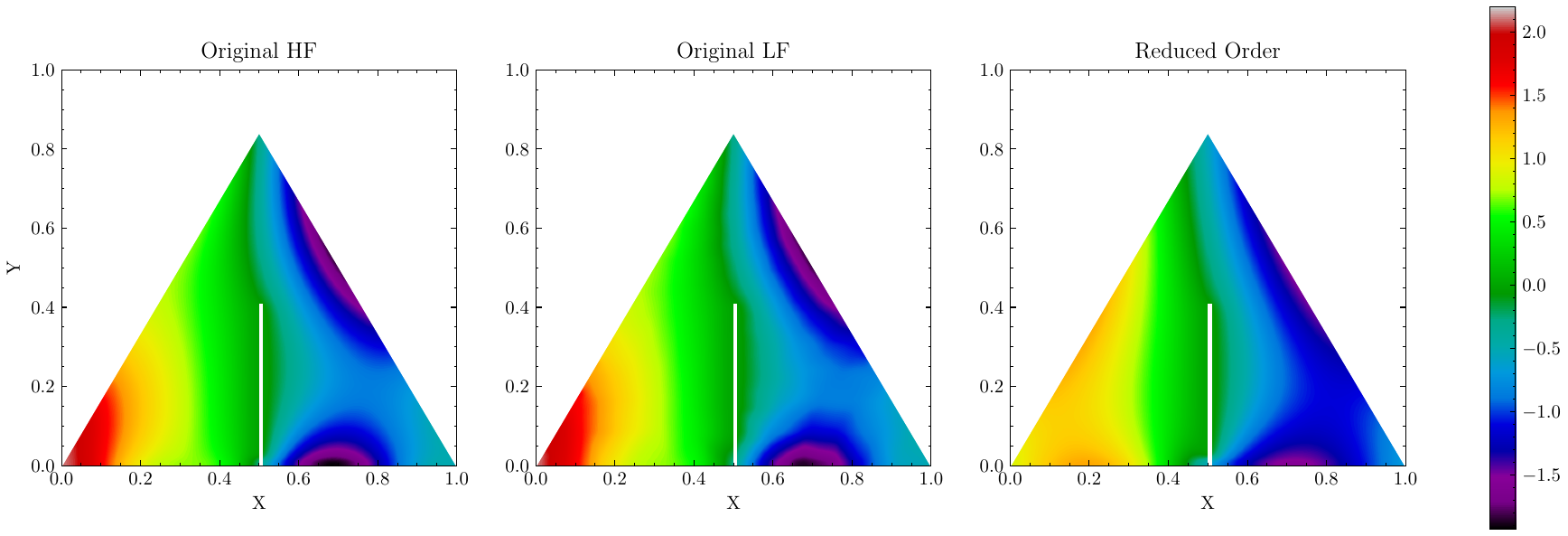}
    \caption{Comparison between the original high-fidelity data, original lower-fidelity data, and the reduced-order lower-fidelity data.}
    \label{fig: 2DRomdata}
\end{figure}

With this approach, we preserve the overall structure of the solution while selectively removing finer details in the lower-fidelity data, mimicking the multi-fidelity scenario based on governing equations, as seen in fluid dynamics problems. This method significantly increases the discrepancy, measured by mean absolute error between the low-fidelity (LF) and high-fidelity (HF) data to 0.151, compared to 0.02647 in the original LF setup. An example realisation is illustrated in Fig. \ref{fig: 2DRomdata}. 

We adopt the same model hyperparameters as the two-dimensional WNO configuration specified in Table~\ref{tab:2D WNO hyperparams}. To ensure robustness, each experiment is repeated with 10 different random splits of training and testing data. The results, summarized in Table~\ref{tab: mod 2D WNO results}, show that transfer learning consistently outperforms the other methods across various combinations of low-fidelity (LF) and high-fidelity (HF) training data. The benefits of incorporating LF data are particularly evident when the HF data is limited. In such scenarios, LF data significantly enhances prediction accuracy. However, when a sufficient amount of HF training data is available, the advantage of incorporating LF data diminishes and, in some cases, may even slightly degrade performance relative to the HF-only baseline. In this test case, we found a more pronounced effect of the benefit of transfer learning compared to the other methods when the LF data is less accurate. 

\begin{figure}[H]
    \centering
    \includegraphics[width=0.8\textwidth]{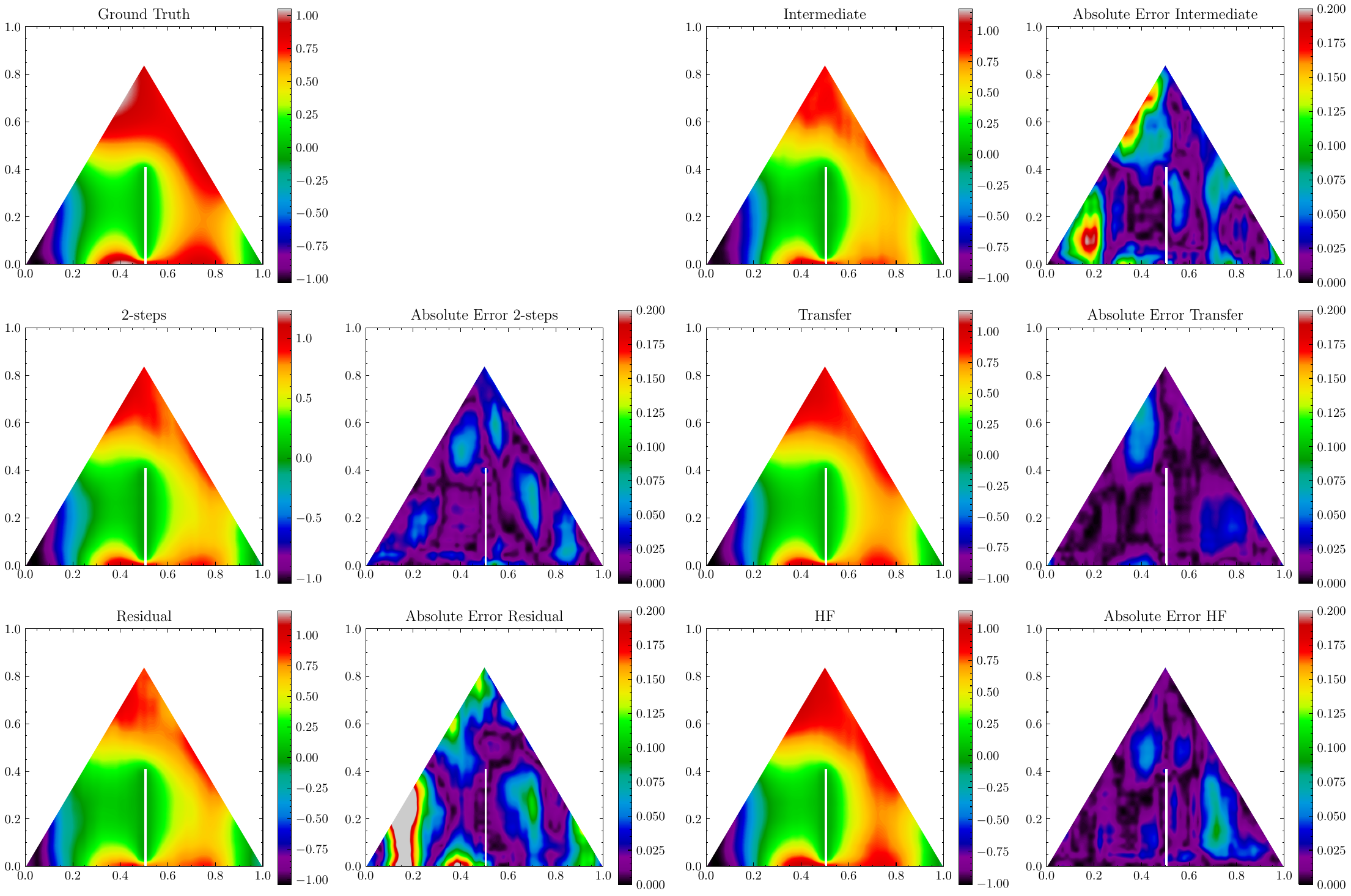}
    \caption{Prediction and error plot for each model on modified 2D Darcy flow in triangular domain with a notch.}
    \label{fig: DarcyROM_realization}
\end{figure}

Figure~\ref{fig: DarcyROM_realization} illustrates a sample realization for a single test case. For both the intermediate and residual models, the largest prediction errors occur near the top-left and bottom-left corners of the domain. This appears to be coincidental, as the overall error patterns are random and do not exhibit any consistent spatial structure.

\begin{table}[!ht]
\caption{2-dimensional modified Darcy flow multi-fidelity WNO RMSE.}
\resizebox{\textwidth}{!}{%
    \begin{tabular}{lllllll}
        \hline
        \multicolumn{2}{l}{$n_{\text{train}}$} & \multirow{2}{*}{2-steps} & \multirow{2}{*}{Residual} & \multirow{2}{*}{Intermediate} & \multirow{2}{*}{Transfer learning} & \multirow{2}{*}{HF baseline} \\ \cline{1-2}
        LF  & HF & & & & \\ \hline
        50  & 50 & $\num{5.35e-2} \pm \num{6.14e-3}$ & $\num{1.54e-1} \pm \num{1.08e-2}$ & $\num{4.52e-2} \pm \num{3.63e-3}$ & $\mathbf{\num{2.76e-2} \pm \num{3.08e-3}}$ & $\num{3.14e-2} \pm \num{4.54e-3}$ \\
        100 & 50 & $\num{5.65e-2} \pm \num{6.74e-3}$ & $\num{1.32e-1} \pm \num{4.87e-3}$ & $\num{6.82e-2} \pm \num{6.58e-3}$ & $\mathbf{\num{2.74e-2} \pm \num{4.38e-3}}$ & $\num{3.14e-2} \pm \num{4.54e-3}$ \\
        150 & 50 & $\num{5.51e-2} \pm \num{8.08e-3}$ & $\num{1.20e-1} \pm \num{7.88e-3}$ & $\num{6.78e-2} \pm \num{9.51e-3}$ & $\mathbf{\num{2.48e-2} \pm \num{1.51e-3}}$ & $\num{3.14e-2} \pm \num{4.54e-3}$ \\
        150 & 25 & $\num{9.47e-2} \pm \num{1.11e-2}$ & $\num{1.59e-1} \pm \num{1.91e-2}$ & $\num{9.26e-2} \pm \num{5.04e-3}$ & $\mathbf{\num{4.63e-2} \pm \num{6.39e-3}}$ & $\num{8.86e-2} \pm \num{1.81e-2}$ \\
        150 & 15 & $\num{1.65e-1} \pm \num{3.16e-2}$ & $\num{3.23e-1} \pm \num{1.51e-1}$ & $\num{1.27e-1} \pm \num{2.57e-2}$ & $\mathbf{\num{7.62e-2} \pm \num{6.65e-3}}$ & $\num{2.27e-1} \pm \num{4.32e-2}$ \\ \hline
    \end{tabular}}
    \label{tab: mod 2D WNO results}
\end{table}

\begin{figure}[H]
    \centering
    \includegraphics[width=0.8\textwidth]{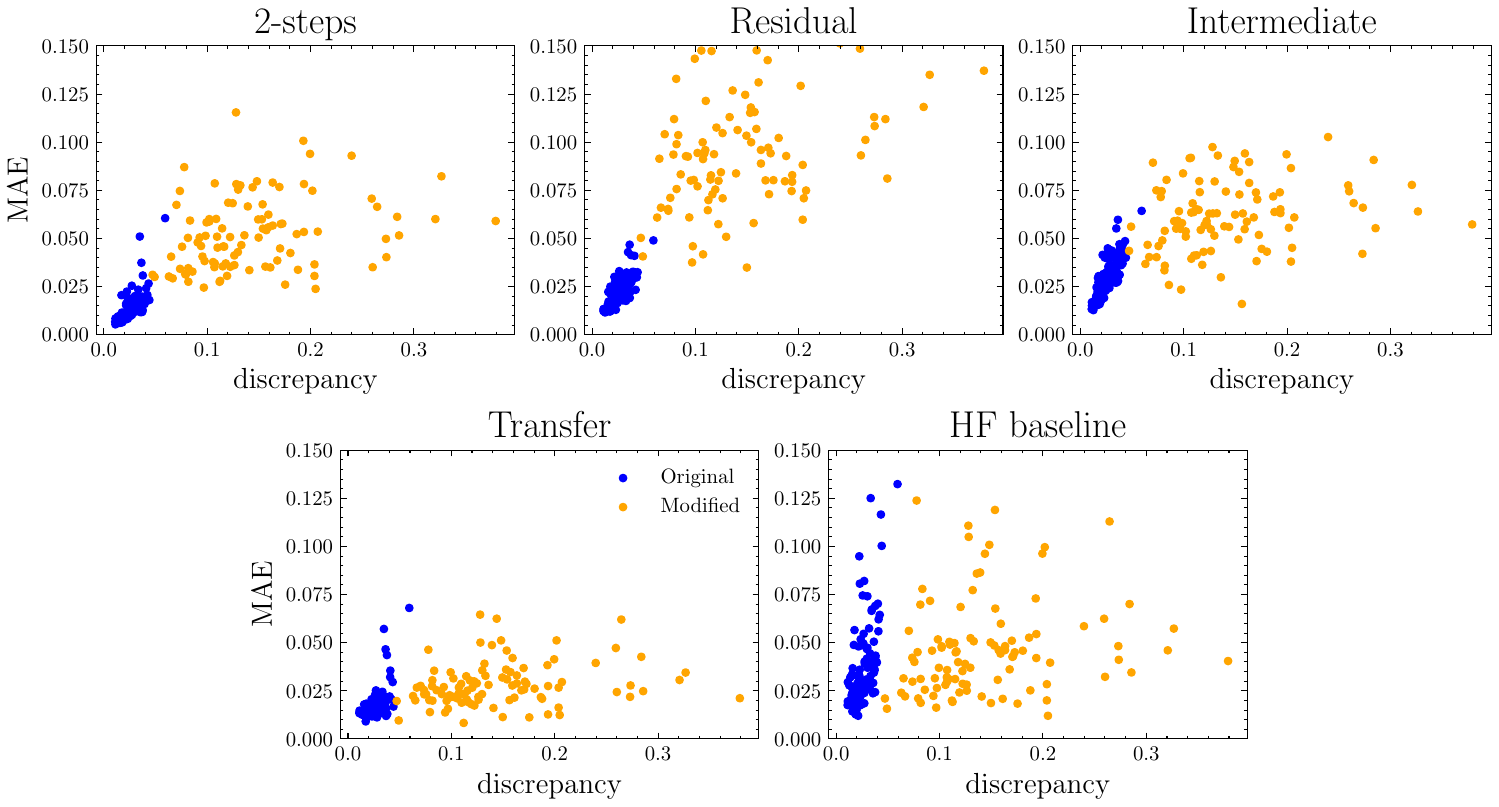}
    \caption{Discrepancy vs mean absolute error (MAE) plot for the 2-dimensional Darcy flow problem. Blue circle indicates the original data points from section \ref{sec: 2ddarcy}, and the orange circle indicates data points from section \ref{sec: mod2ddarcy}.}
    \label{fig: DarcyROM_discrepancy}
\end{figure}

To further analyse the models' performance under varying levels of discrepancy between low-fidelity (LF) and high-fidelity (HF) data, we examine the relationship between the LF-HF discrepancy and the mean absolute error (MAE) of the predictions compared to the HF ground truth. As illustrated in Fig. \ref{fig: DarcyROM_discrepancy}, prediction errors generally increase with larger LF-HF discrepancies, a trend that is expected. However, the extent to which the discrepancy affects model performance varies. Notably, the transfer learning approach appears to be the least sensitive to increasing discrepancies. While we do not provide a theoretical justification, we hypothesise that this robustness may stem from the way LF information is incorporated. In the intermediate, residual, and two-step models, LF data is directly injected into the input of the HF model, which may confuse the network when the discrepancy is large. In contrast, transfer learning leverages LF-trained weights as initialisation for the HF model. This approach allows the HF model to start with a rough understanding of the solution’s general structure, which is then refined during training on HF data, making it more resilient to LF-HF mismatch.

\begin{figure}[H]
    \centering
    \includegraphics[width=0.8\textwidth]{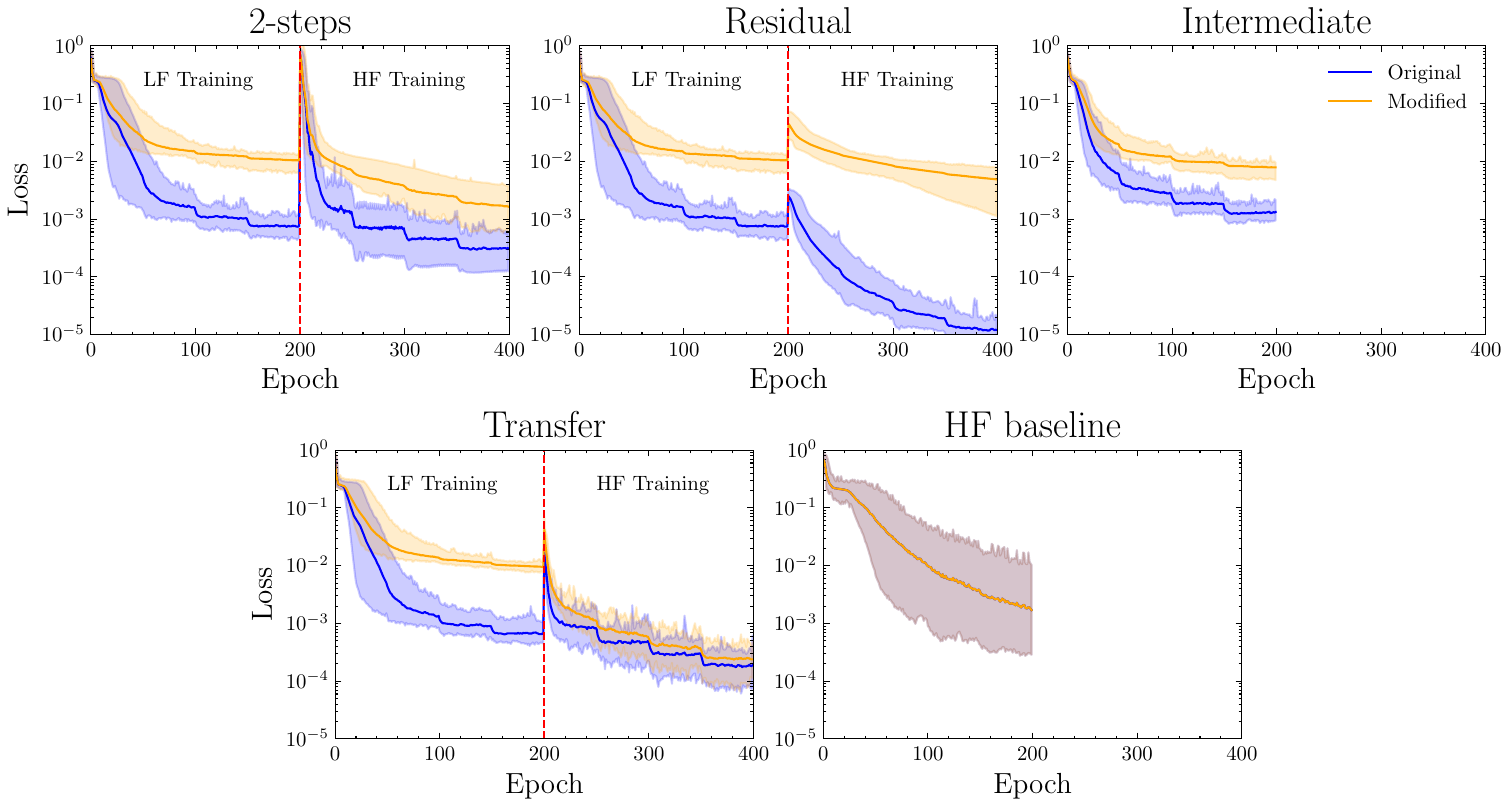}
    \caption{Training history for Darcy flow problem. Blue lines represent training histories using the original dataset described in Section \ref{sec: 2ddarcy}, while orange lines correspond to training histories using the modified dataset from Section \ref{sec: mod2ddarcy}. }
    \label{fig: Darcy_history}
\end{figure}

We also investigate how different models perform under varying levels of discrepancy between LF and HF data. The first scenario, presented in Section \ref{sec: 2ddarcy}, features relatively low LF-HF discrepancy, while the second scenario in Section \ref{sec: mod2ddarcy} introduces a higher discrepancy by discarding more detailed information in the LF data. As shown in Fig.~\ref{fig: Darcy_history}, the LF training losses exhibit a clear gap between the original and modified datasets, an expected outcome due to the reduced complexity in the modified LF data. Interestingly, the differences become more pronounced during the HF training phase. The residual model performs well on the original dataset but struggles significantly on the modified data, suggesting its sensitivity to LF-HF mismatch. In contrast, the transfer learning model demonstrates strong performance across both scenarios, as indicated by the similar final loss values. The two-step model lies between the two; it performs better than the residual model on the modified data, but does not match the robustness of transfer learning. For the intermediate model, which jointly trains on LF and HF data in a single pass, training loss is plotted only up to epoch 200. All models are trained for a total of 200 epochs. For those employing separate LF and HF phases (e.g., two-step, residual, and transfer learning models), HF training is plotted immediately following LF training for visualisation purposes, although in practice, these phases may occur separately. Lastly, the HF baseline yields identical training curves for both datasets, as the HF data is unchanged.

\subsection{Unsteady smoke inflow}
The multi-fidelity unsteady smoke inflow dataset, introduced in \cite{ZakariaSmoke2024}, models the time-dependent concentration of smoke particles within a rectangular domain. It consists of 125 unique inlet configurations, each simulated over 126 timesteps. The inlet has a fixed radius of 3 units, and its location is sampled using Latin Hypercube Sampling (LHS), constrained within the ranges $[4, 28]$ along the horizontal axis and $[4, 35]$ along the vertical axis. The objective is to learn the temporal evolution of smoke particle concentration for varying inlet positions. To this end, the model is trained on data from different inlet configurations across all timesteps and tested by predicting the concentration evolution over the first 50 timesteps for the unseen inlet locations. The smoke inflow problem itself is driven by the Navier-Stokes equation, defined as:
\begin{equation}
    \begin{gathered}
        \frac{\partial \boldsymbol{u}}{\partial t} = - (\boldsymbol{u} \cdot \boldsymbol{\nabla}) \boldsymbol{u} - \nu \boldsymbol{\nabla}^2 \boldsymbol{u} - \frac 1 \rho \boldsymbol{\nabla} p + g, \\
        \frac{\partial s}{\partial t} + (\boldsymbol{u} \cdot \boldsymbol{\nabla})s = \alpha \boldsymbol{\nabla}^2s + i.
    \end{gathered}
    \label{eq: NavierStokes}
\end{equation}
Where $\boldsymbol{u}$ is the velocity component, $\rho$ and $p$ denote the fluid density and pressure, $\alpha$ denotes the particle diffusivity, $s$ and $i$ are the smoke particle concentration and the particle concentration inflow, respectively. The flow simulations were conducted using PhiFlow \cite{holl2024phiflow} package, where the simulations are set to be incompressible, viscous, and laminar. The authors follow the standard simulation settings with viscosity $\nu=0.1$ and density $\rho=1.0$, described in the PhiFlow smoke plume tutorial.\footnote{Available at: \url{https://tum-pbs.github.io/PhiFlow/examples/grids/Smoke_Plume.html}, accessed May 15, 2025.}

\begin{figure}[H]
    \centering
    \includegraphics[width=0.65\textwidth]{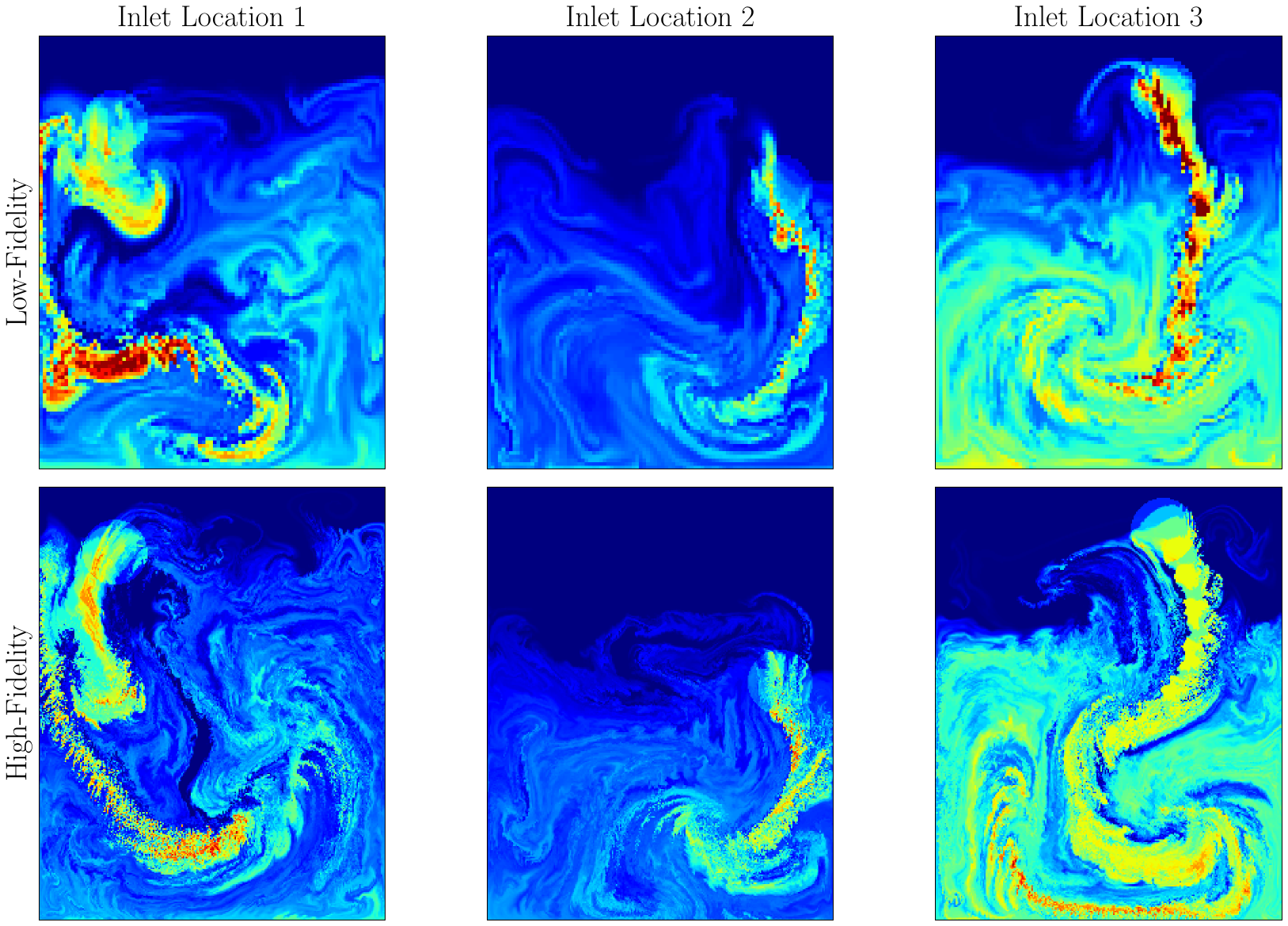}
    \caption{Low and high-fidelity smoke inflow data with different inlet locations at timestep $t=75$.}
    \label{fig: smokedata}
\end{figure}

The low-fidelity data is represented on a $96 \times 120$ grid, while the high-fidelity data uses a finer $256 \times 320$ grid. Although fidelity levels were controlled by grid resolution, the discrepancy between LF and HF data is notably larger than in the 2D Darcy flow, as shown in Fig. \ref{fig: 2DRomdata} between the Original HF and LF. The unsteady smoke inflow problem, illustrated in Fig. \ref{fig: smokedata}, shows that the LF simulations are able to capture the general trend of the flow, but somehow fail to predict the exact value.

Given the high computational cost to train the model, our experiments were limited to 100 low-fidelity (LF) and 10 high-fidelity (HF) training samples, with each experiment repeated five times. Since the problem is time-dependent, the number of rollouts during training plays a crucial role \cite{lippe2023pderefiner,kohl2024autoreg}. However, constrained by hardware limitations, we restrict the rollout to a single step and use only a subset of the time domain, from $t=0$ to $t=50$, rather than using full timesteps for training. The remaining hyperparameters for the LF and HF models are listed in Table \ref{tab:smoke FNO}.

\begin{table}[!ht]
    \centering
    \caption{Smoke inflow FNO models hyperparameters.}
    \begin{tabular}{l l l}
        \hline
         & Lower-fidelity & Higher-fidelity \\
        \hline
        NO kernel & Fourier & Fourier \\
        Number of NO layers & 2 & 2 \\
        FC layers width & 32 & 32 \\
        Fourier modes & 16 & 16 \\
        Rollout number & 3 & 3 \\
        Learning rate& $4\times 10^{-4}$ & $5\times 10^{-4}$ \\
        Weight Decay & $1\times 10^{-3}$ & $1\times 10^{-2}$ \\
        Epochs & 200 & 200 \\
        \hline
    \end{tabular}
    \label{tab:smoke FNO}
\end{table}

The model was trained on a single NVIDIA A100 GPU. However, due to the problem’s complexity, we must limit certain training hyperparameters, particularly for the HF model, such as the rollout number and batch size. Specifically, the HF model is trained with a rollout number of one and a maximum batch size of 25. To keep the model size manageable for the available hardware, the fully connected layer width and the number of Fourier modes were limited to 32 and 16, respectively. For the same reason, we did not evaluate the intermediate model in this test case, as it integrates both LF and HF inputs, requiring a larger batch size that exceeds our hardware constraints. Also, since the 2-step model performs better in most 2D test cases than the residual model, we only compare the 2-step model, transfer learning, and the HF baseline.

\begin{figure}[h]
    \centering
    \includegraphics[width=0.5\textwidth]{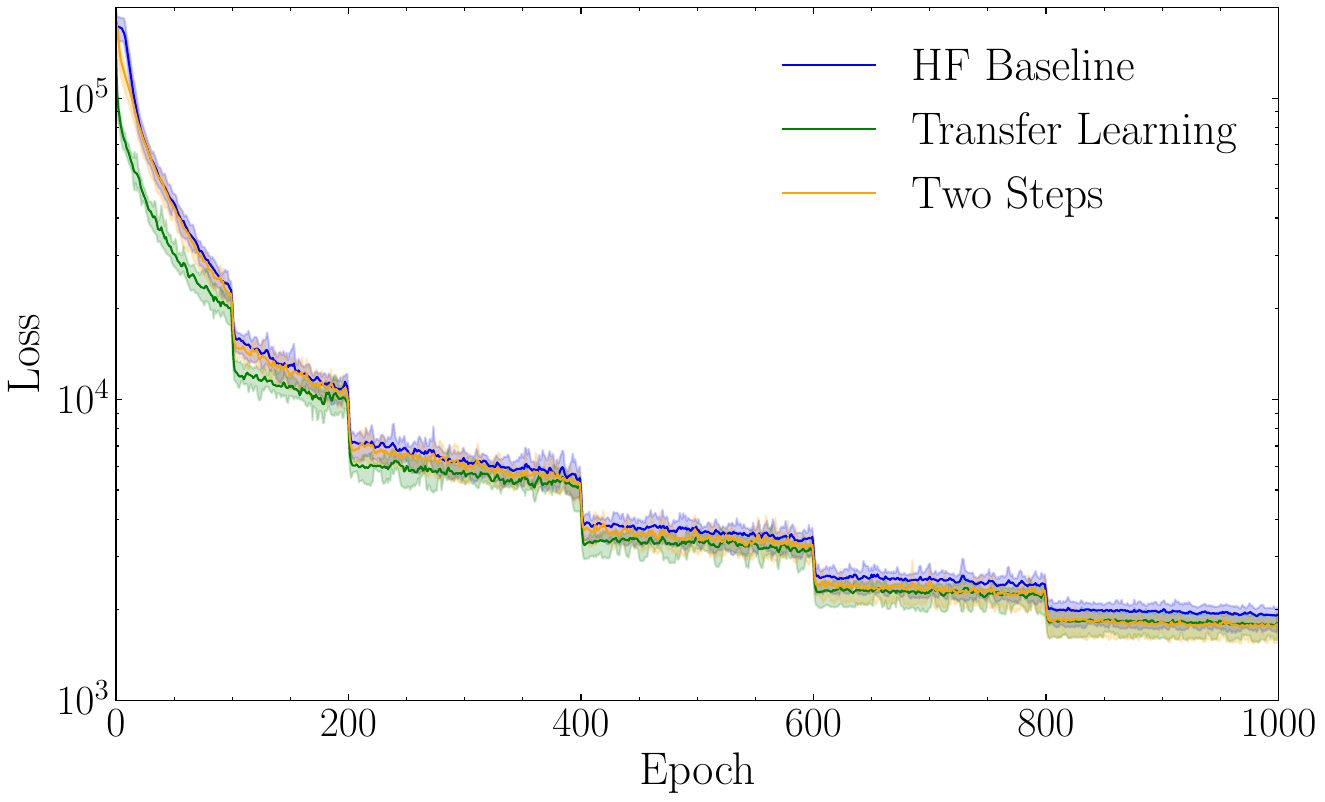}
    \caption{HF training phase loss history for smoke inflow problem.}
    \label{fig: smoke_hist}
\end{figure}

During training, a single LF model is shared across all multi-fidelity approaches to ensure a fair comparison. Consequently, the LF model's loss history is not included in Fig.~\ref{fig: smoke_hist}. The HF training phase is repeated five times with different random seeds and training sample selections to assess consistency. As illustrated in Fig.~\ref{fig: smoke_hist}, the resulting loss values across models show no significant variation, indicating that all models are capable of effectively learning from the data. 

\begin{figure}[h]
    \centering
    \includegraphics[width=0.5\textwidth]{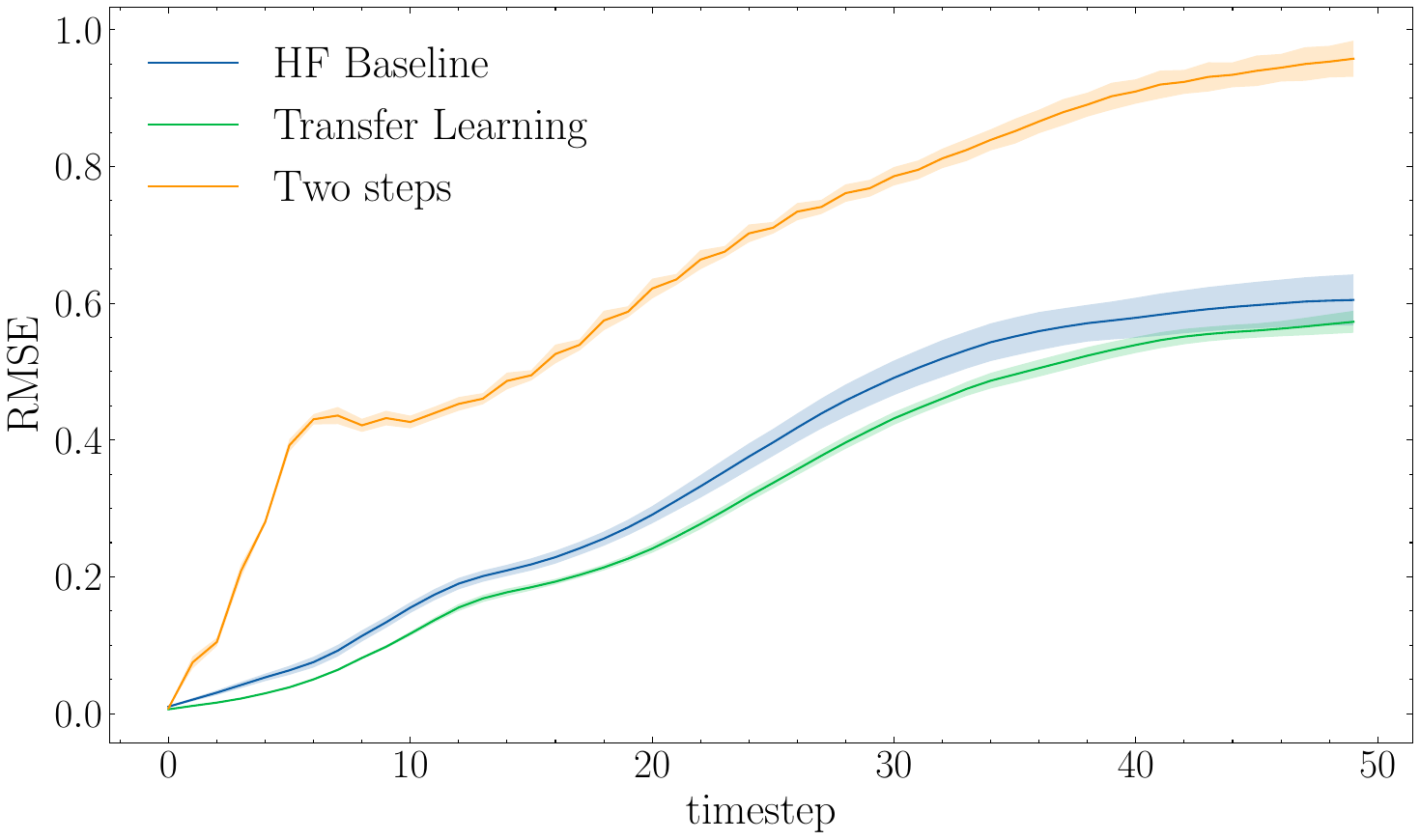}
    \caption{RMSE at each timestep for each multi-fidelity strategy.}
    \label{fig: smokermse}
\end{figure}

The model is evaluated on 25 unseen inlet locations using the high-fidelity (HF) dataset. Unlike the training loss history, the evaluation results, shown in Fig.~\ref{fig: smokermse}, demonstrate that the transfer learning approach consistently outperforms the HF baseline across all timesteps. In contrast, while having a similar profile for training loss, the 2-step model fails to generalise to unseen data, indicated by larger numbers of errors. This suggests that incorporating the LF model output as an additional input feature in the HF model tends to degrade prediction performance rather than enhance it.

\begin{figure}[h]
    \centering
    \includegraphics[width=0.5\textwidth]{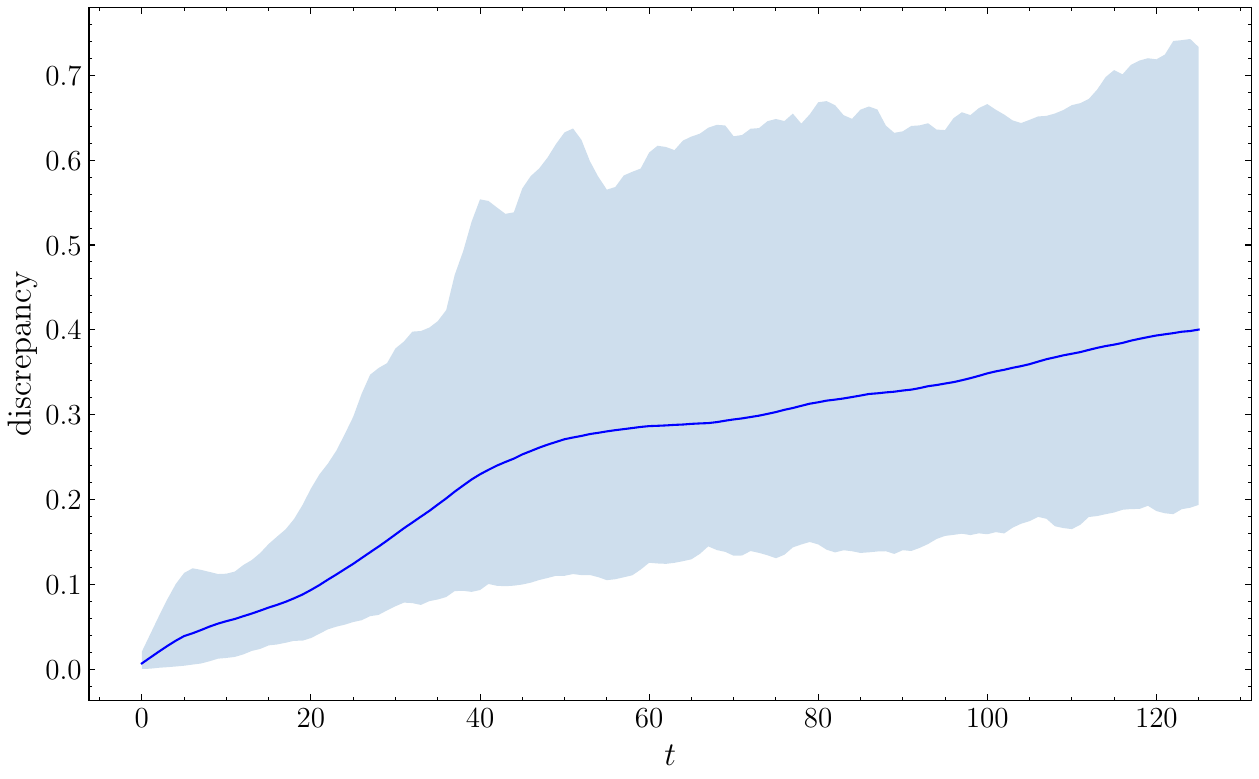}
    \caption{Data discrepancy vs time for the smoke inflow data.}
    \label{fig: smokedisc}
\end{figure}

Given the temporal nature of the data, we further analyse how discrepancies between LF and HF evolve over time to better understand the models' behavior. In Fig.~\ref{fig: smokedisc}, we present the discrepancy computed across all inlet locations in the dataset at each time step. The figure indicates that the discrepancy generally increases as time progresses. However, this increasing divergence is not the sole challenge. Due to the sequential nature of the problem, both the training and inference phases rely on an autoregressive strategy, where predictions from previous time steps are used as inputs for subsequent steps. This setup can lead to error accumulation over time. For multi-fidelity models that are particularly sensitive to discrepancies, such as the two-step model, this compounding effect can significantly degrade performance.

\section{Conclusion}
\label{sec:conclusion}
As PDE simulation problems grow in complexity, generating high-fidelity (HF) data for training surrogate models like Neural Operators (NOs) becomes increasingly costly. To mitigate this challenge, multi-fidelity learning has emerged as a promising strategy to leverage both low- and high-fidelity data during training. In this paper, we investigate several multi-fidelity approaches for NOs, including transfer learning, a two-step model, a residual-enhanced two-step model, and an intermediate architecture. These methods are evaluated against a baseline NO trained solely on HF data, using a series of test cases with increasing complexity. Our results show that while most multi-fidelity strategies perform comparably to the HF baseline on simpler tasks, their benefits diminish as the problem becomes more challenging. Notably, among all methods, only the transfer learning approach consistently outperforms the HF baseline across all scenarios.

Further analysis under varying LF-HF discrepancies revealed that strategies incorporating LF information directly into the HF model input are more vulnerable to performance degradation, especially when discrepancies are large. This vulnerability likely arises from the conflicting information that hinders effective learning. In contrast, transfer learning, by using LF-trained weights to initialise the HF model, provides a better learning process. It offers a generalised prior that is refined through HF training, making it more robust to LF-HF mismatch. This resilience becomes especially critical in complex, temporally unsteady problems where an autoregressive scheme can accumulates errors over time. In such settings, direct LF incorporation strategies struggle, whereas transfer learning maintains stable and reliable performance.

In addition to model exploration, we introduce two test cases situated at the higher end of the complexity spectrum. These were motivated by the current scarcity of complex multi-fidelity benchmarks. We observed that many existing multi-fidelity studies primarily focus on fidelity differences arising from grid resolution. While such cases are valid, they capture only a narrow subset of the broader multi-fidelity learning landscape. To address this, we propose a modification to the existing two-dimensional Darcy flow dataset by systematically reducing the physical complexity of the low-fidelity data using proper orthogonal decomposition (POD). Additionally, we introduce a new test case involving two-dimensional unsteady laminar smoke inflow, designed to evaluate the models' ability to capture dynamic, time-evolving phenomena.

While this paper demonstrates the potential of multi-fidelity modeling for unsteady flow problems, the topic remains a rich and promising avenue for future research. One important direction is to investigate more effective implementations of multi-fidelity strategies, particularly in complex temporal problems such as the smoke inflow scenario. For instance, understanding why certain strategies fail to generalize to unseen data warrants further analysis. Exploring alternative neural operator architectures could also provide valuable insights and improved performance. Additionally, the current smoke inflow test case does not reflect fidelity differences arising from variations in governing equations. Developing new benchmark cases that incorporate such discrepancies would be a valuable contribution. Another promising direction is to explore uncertainty quantification in both low-fidelity and high-fidelity models, specifically, how different sources of uncertainty interact and propagate through multi-fidelity frameworks. 

\section*{Acknowledgments}
This work was supported by granted H2020 FETOPEN-2018-2019-2020-01 European project, \emph{Epistemic AI} under grant agreement No. 964505 (E-pi).



 \bibliographystyle{elsarticle-num} 
 \bibliography{cas-refs}

@article{Guo2022,
  title = {Multi-fidelity regression using artificial neural networks: Efficient approximation of parameter-dependent output quantities},
  volume = {389},
  ISSN = {0045-7825},
  url = {http://dx.doi.org/10.1016/j.cma.2021.114378},
  DOI = {10.1016/j.cma.2021.114378},
  journal = {Computer Methods in Applied Mechanics and Engineering},
  publisher = {Elsevier BV},
  author = {Guo,  Mengwu and Manzoni,  Andrea and Amendt,  Maurice and Conti,  Paolo and Hesthaven,  Jan S.},
  year = {2022},
  month = {feb},
  pages = {114378}
}

@article{Tripura2024,
  title = {Multi-fidelity wavelet neural operator surrogate model for time-independent and time-dependent reliability analysis},
  volume = {77},
  ISSN = {0266-8920},
  url = {http://dx.doi.org/10.1016/j.probengmech.2024.103672},
  DOI = {10.1016/j.probengmech.2024.103672},
  journal = {Probabilistic Engineering Mechanics},
  publisher = {Elsevier BV},
  author = {Tripura,  Tapas and Thakur,  Akshay and Chakraborty,  Souvik},
  year = {2024},
  month = {jul},
  pages = {103672}
}

@inproceedings{Drucker1996,
 author = {Drucker, Harris and Burges, Christopher J. C. and Kaufman, Linda and Smola, Alex and Vapnik, Vladimir},
 booktitle = {Advances in Neural Information Processing Systems},
 editor = {M.C. Mozer and M. Jordan and T. Petsche},
 pages = {1},
 publisher = {MIT Press},
 title = {Support Vector Regression Machines},
 volume = {9},
 year = {1996}
}

@article{tripathy2018deep,
  title={Deep UQ: Learning deep neural network surrogate models for high dimensional uncertainty quantification},
  author={Tripathy, Rohit K and Bilionis, Ilias},
  journal={Journal of computational physics},
  volume={375},
  pages={565--588},
  year={2018},
  publisher={Elsevier}
}

@article{sun2019review,
  title={A review of the artificial neural network surrogate modeling in aerodynamic design},
  author={Sun, Gang and Wang, Shuyue},
  journal={Proceedings of the Institution of Mechanical Engineers, Part G: Journal of Aerospace Engineering},
  volume={233},
  number={16},
  pages={5863--5872},
  year={2019},
  publisher={SAGE Publications Sage UK: London, England}
}

@BOOK{Rasmussen2005-at,
  title     = "Gaussian processes for machine learning",
  author    = "Rasmussen, Carl Edward and Williams, Christopher K I",
  publisher = "MIT Press",
  series    = "Adaptive Computation and Machine Learning series",
  month     =  "nov",
  year      =  "2005",
  address   = "London, England",
  language  = "en"
}

@article{Bernardini2021,
  title = {STREAmS: A high-fidelity accelerated solver for direct numerical simulation of compressible turbulent flows},
  volume = {263},
  ISSN = {0010-4655},
  url = {http://dx.doi.org/10.1016/j.cpc.2021.107906},
  DOI = {10.1016/j.cpc.2021.107906},
  journal = {Computer Physics Communications},
  publisher = {Elsevier BV},
  author = {Bernardini,  Matteo and Modesti,  Davide and Salvadore,  Francesco and Pirozzoli,  Sergio},
  year = {2021},
  month = {jun},
  pages = {107906}
}

@article{Huthwaite2014,
  title = {Accelerated finite element elastodynamic simulations using the GPU},
  volume = {257},
  ISSN = {0021-9991},
  url = {http://dx.doi.org/10.1016/j.jcp.2013.10.017},
  DOI = {10.1016/j.jcp.2013.10.017},
  journal = {Journal of Computational Physics},
  publisher = {Elsevier BV},
  author = {Huthwaite,  Peter},
  year = {2014},
  month = {jan},
  pages = {687–707}
}

@inproceedings{renard2021,
  TITLE = {{ZONAL DETACHED EDDY SIMULATION OF UNSTEADY AIRFOIL AERODYNAMICS}},
  AUTHOR = {Renard, Nicolas and Deck, Sebastien},
  URL = {https://hal.science/hal-03206330},
  BOOKTITLE = {{AERO 2020+1 - 55th 3AF International Conference on Applied Conference}},
  ADDRESS = {Poitiers (virtuel), France},
  YEAR = {2021},
  MONTH = {Apr},
  HAL_ID = {hal-03206330},
  HAL_VERSION = {v1},
  pages = {1}
}

@misc{li2021fno,
      title={Fourier Neural Operator for Parametric Partial Differential Equations}, 
      author={Zongyi Li and Nikola Kovachki and Kamyar Azizzadenesheli and Burigede Liu and Kaushik Bhattacharya and Andrew Stuart and Anima Anandkumar},
      year={2021},
      eprint={2010.08895},
      archivePrefix={arXiv},
      primaryClass={cs.LG},
      url={https://arxiv.org/abs/2010.08895} 
}

@misc{Kovachki2021,
  doi = {10.48550/ARXIV.2108.08481},
  url = {https://arxiv.org/abs/2108.08481},
  author = {Kovachki,  Nikola and Li,  Zongyi and Liu,  Burigede and Azizzadenesheli,  Kamyar and Bhattacharya,  Kaushik and Stuart,  Andrew and Anandkumar,  Anima},
  title = {Neural Operator: Learning Maps Between Function Spaces},
  publisher = {arXiv},
  year = {2021},
  copyright = {arXiv.org perpetual,  non-exclusive license}
}

@misc{li2020gno,
      title={Neural Operator: Graph Kernel Network for Partial Differential Equations}, 
      author={Zongyi Li and Nikola Kovachki and Kamyar Azizzadenesheli and Burigede Liu and Kaushik Bhattacharya and Andrew Stuart and Anima Anandkumar},
      year={2020},
      eprint={2003.03485},
      archivePrefix={arXiv},
      primaryClass={cs.LG},
      url={https://arxiv.org/abs/2003.03485}
}

@misc{hendrycks2023gelu,
      title={Gaussian Error Linear Units (GELUs)}, 
      author={Dan Hendrycks and Kevin Gimpel},
      year={2023},
      eprint={1606.08415},
      archivePrefix={arXiv},
      primaryClass={cs.LG},
      url={https://arxiv.org/abs/1606.08415} 
}

@misc{elfwing2017silu,
      title={Sigmoid-Weighted Linear Units for Neural Network Function Approximation in Reinforcement Learning}, 
      author={Stefan Elfwing and Eiji Uchibe and Kenji Doya},
      year={2017},
      eprint={1702.03118},
      archivePrefix={arXiv},
      primaryClass={cs.LG},
      url={https://arxiv.org/abs/1702.03118} 
}

@misc{agarap2019relu,
      title={Deep Learning using Rectified Linear Units (ReLU)}, 
      author={Abien Fred Agarap},
      year={2019},
      eprint={1803.08375},
      archivePrefix={arXiv},
      primaryClass={cs.NE},
      url={https://arxiv.org/abs/1803.08375}
}

@article{Meng2020,
  title = {A composite neural network that learns from multi-fidelity data: Application to function approximation and inverse PDE problems},
  volume = {401},
  ISSN = {0021-9991},
  url = {http://dx.doi.org/10.1016/j.jcp.2019.109020},
  DOI = {10.1016/j.jcp.2019.109020},
  journal = {Journal of Computational Physics},
  publisher = {Elsevier BV},
  author = {Meng,  Xuhui and Karniadakis,  George Em},
  year = {2020},
  month = {jan},
  pages = {109020}
}

@article{Forrester2007,
  title = {Multi-fidelity optimization via surrogate modelling},
  volume = {463},
  ISSN = {1471-2946},
  url = {http://dx.doi.org/10.1098/rspa.2007.1900},
  DOI = {10.1098/rspa.2007.1900},
  number = {2088},
  journal = {Proceedings of the Royal Society A: Mathematical,  Physical and Engineering Sciences},
  publisher = {The Royal Society},
  author = {Forrester,  Alexander I.J and Sóbester,  András and Keane,  Andy J},
  year = {2007},
  month = {oct},
  pages = {3251–3269}
}

@article{Palar2016,
  title = {Multi-fidelity non-intrusive polynomial chaos based on regression},
  volume = {305},
  ISSN = {0045-7825},
  url = {http://dx.doi.org/10.1016/j.cma.2016.03.022},
  DOI = {10.1016/j.cma.2016.03.022},
  journal = {Computer Methods in Applied Mechanics and Engineering},
  publisher = {Elsevier BV},
  author = {Palar,  Pramudita Satria and Tsuchiya,  Takeshi and Parks,  Geoffrey Thomas},
  year = {2016},
  month = {jun},
  pages = {579–606}
}

@article{Chakraborty2021,
  title = {Transfer learning based multi-fidelity physics informed deep neural network},
  volume = {426},
  ISSN = {0021-9991},
  url = {http://dx.doi.org/10.1016/j.jcp.2020.109942},
  DOI = {10.1016/j.jcp.2020.109942},
  journal = {Journal of Computational Physics},
  publisher = {Elsevier BV},
  author = {Chakraborty,  Souvik},
  year = {2021},
  month = {feb},
  pages = {109942}
}

@article{Lyu_2023,
   title={Multi-fidelity prediction of fluid flow based on transfer learning using Fourier neural operator},
   volume={35},
   ISSN={1089-7666},
   url={http://dx.doi.org/10.1063/5.0155555},
   DOI={10.1063/5.0155555},
   number={7},
   journal={Physics of Fluids},
   publisher={AIP Publishing},
   author={Lyu, Yanfang and Zhao, Xiaoyu and Gong, Zhiqiang and Kang, Xiao and Yao, Wen},
   year={2023},
   month={jul} }

@misc{tang2024,
      title={Multi-fidelity Fourier Neural Operator for Fast Modeling of Large-Scale Geological Carbon Storage}, 
      author={Hewei Tang and Qingkai Kong and Joseph P. Morris},
      year={2024},
      eprint={2308.09113},
      archivePrefix={arXiv},
      primaryClass={stat.ML},
      url={https://arxiv.org/abs/2308.09113}
}

@article{Koziel2014,
  title = {Efficient Multi-Objective Simulation-Driven Antenna Design Using Co-Kriging},
  volume = {62},
  ISSN = {1558-2221},
  url = {http://dx.doi.org/10.1109/TAP.2014.2354673},
  DOI = {10.1109/tap.2014.2354673},
  number = {11},
  journal = {IEEE Transactions on Antennas and Propagation},
  publisher = {Institute of Electrical and Electronics Engineers (IEEE)},
  author = {Koziel,  Slawomir and Bekasiewicz,  Adrian and Couckuyt,  Ivo and Dhaene,  Tom},
  year = {2014},
  month = {nov},
  pages = {5900–5905}
}

@article{Liu2022,
  title = {Multi-fidelity Co-Kriging surrogate model for ship hull form optimization},
  volume = {243},
  ISSN = {0029-8018},
  url = {http://dx.doi.org/10.1016/j.oceaneng.2021.110239},
  DOI = {10.1016/j.oceaneng.2021.110239},
  journal = {Ocean Engineering},
  publisher = {Elsevier BV},
  author = {Liu,  Xinwang and Zhao,  Weiwen and Wan,  Decheng},
  year = {2022},
  month = {jan},
  pages = {110239}
}

@article{FernndezGodino2019,
  title = {Linear regression-based multifidelity surrogate for disturbance amplification in multiphase explosion},
  volume = {60},
  ISSN = {1615-1488},
  url = {http://dx.doi.org/10.1007/s00158-019-02387-4},
  DOI = {10.1007/s00158-019-02387-4},
  number = {6},
  journal = {Structural and Multidisciplinary Optimization},
  publisher = {Springer Science and Business Media LLC},
  author = {Fernández-Godino,  M. Giselle and Dubreuil,  Sylvain and Bartoli,  Nathalie and Gogu,  Christian and Balachandar,  S. and Haftka,  Raphael T.},
  year = {2019},
  month = {oct},
  pages = {2205–2220}
}

@article{Zhang2018,
  title = {Multifidelity Surrogate Based on Single Linear Regression},
  volume = {56},
  ISSN = {1533-385X},
  url = {http://dx.doi.org/10.2514/1.J057299},
  DOI = {10.2514/1.j057299},
  number = {12},
  journal = {AIAA Journal},
  publisher = {American Institute of Aeronautics and Astronautics (AIAA)},
  author = {Zhang,  Yiming and Kim,  Nam H. and Park,  Chanyoung and Haftka,  Raphael T.},
  year = {2018},
  month = {dec},
  pages = {4944–4952}
}

@article{GiselleFernndezGodino2023,
  title = {Review of multi-fidelity models},
  volume = {1},
  ISSN = {2837-1739},
  url = {http://dx.doi.org/10.3934/acse.2023015},
  DOI = {10.3934/acse.2023015},
  number = {4},
  journal = {Advances in Computational Science and Engineering},
  publisher = {American Institute of Mathematical Sciences (AIMS)},
  author = {Giselle Fernández-Godino,  M.},
  year = {2023},
  pages = {351–400}
}

@article{Myers1982,
  title = {Matrix formulation of co-kriging},
  volume = {14},
  ISSN = {1573-8868},
  url = {http://dx.doi.org/10.1007/BF01032887},
  DOI = {10.1007/bf01032887},
  number = {3},
  journal = {Journal of the International Association for Mathematical Geology},
  publisher = {Springer Science and Business Media LLC},
  author = {Myers,  Donald E.},
  year = {1982},
  month = {jun},
  pages = {249–257}
}

@inbook{Eldred2004,
author = {Michael Eldred and Anthony Giunta and S. Collis},
title = {Second-Order Corrections for Surrogate-Based Optimization with Model Hierarchies},
booktitle = {10th AIAA/ISSMO Multidisciplinary Analysis and Optimization Conference},
chapter = {1},
pages = {1},
doi = {10.2514/6.2004-4457},
URL = {https://arc.aiaa.org/doi/abs/10.2514/6.2004-4457},
eprint = {https://arc.aiaa.org/doi/pdf/10.2514/6.2004-4457},
publisher = {American Institute of Aeronautics and Astronautics},
 year = {2004},
  month = {aug}
}

@article{Robinson2008,
  title = {Surrogate-Based Optimization Using Multifidelity Models with Variable Parameterization and Corrected Space Mapping},
  volume = {46},
  ISSN = {1533-385X},
  url = {http://dx.doi.org/10.2514/1.36043},
  DOI = {10.2514/1.36043},
  number = {11},
  journal = {AIAA Journal},
  publisher = {American Institute of Aeronautics and Astronautics (AIAA)},
  author = {Robinson,  T. D. and Eldred,  M. S. and Willcox,  K. E. and Haimes,  R.},
  year = {2008},
  month = {nov},
  pages = {2814–2822}
}

@article{Leifsson2010,
  title = {Multi-fidelity design optimization of transonic airfoils using physics-based surrogate modeling and shape-preserving response prediction},
  volume = {1},
  ISSN = {1877-7503},
  url = {http://dx.doi.org/10.1016/j.jocs.2010.03.007},
  DOI = {10.1016/j.jocs.2010.03.007},
  number = {2},
  journal = {Journal of Computational Science},
  publisher = {Elsevier BV},
  author = {Leifsson,  Leifur and Koziel,  Slawomir},
  year = {2010},
  month = {jun},
  pages = {98–106}
}

@article{chatterjee_pod,
 ISSN = {00113891},
 URL = {http://www.jstor.org/stable/24103957},
 author = {Anindya Chatterjee},
 journal = {Current Science},
 number = {7},
 pages = {808--817},
 publisher = {Temporary Publisher},
 title = {An introduction to the proper orthogonal decomposition},
 urldate = {2023-05-08},
 volume = {78},
 year = {2000}
}

@article{Murtagh1991,
  title = {Multilayer perceptrons for classification and regression},
  volume = {2},
  ISSN = {0925-2312},
  url = {http://dx.doi.org/10.1016/0925-2312(91)90023-5},
  DOI = {10.1016/0925-2312(91)90023-5},
  number = {5–6},
  journal = {Neurocomputing},
  publisher = {Elsevier BV},
  author = {Murtagh,  Fionn},
  year = {1991},
  month = {jul},
  pages = {183–197}
}

@article{Popescu2009,
author = {Popescu, Marius-Constantin and Balas, Valentina E. and Perescu-Popescu, Liliana and Mastorakis, Nikos},
title = {Multilayer perceptron and neural networks},
year = {2009},
issue_date = {July 2009},
publisher = {World Scientific and Engineering Academy and Society (WSEAS)},
address = {Stevens Point, Wisconsin, USA},
volume = {8},
number = {7},
issn = {1109-2734},
journal = {WSEAS Trans. Cir. and Sys.},
month = {jul},
pages = {579–588},
numpages = {10}
}

@article{Lecun1998,
  title = {Gradient-based learning applied to document recognition},
  volume = {86},
  ISSN = {0018-9219},
  url = {http://dx.doi.org/10.1109/5.726791},
  DOI = {10.1109/5.726791},
  number = {11},
  journal = {Proceedings of the IEEE},
  publisher = {Institute of Electrical and Electronics Engineers (IEEE)},
  author = {Lecun,  Y. and Bottou,  L. and Bengio,  Y. and Haffner,  P.},
  year = {1998},
  pages = {2278–2324}
}

@article{Lu2021,
  title = {Learning nonlinear operators via DeepONet based on the universal approximation theorem of operators},
  volume = {3},
  ISSN = {2522-5839},
  url = {http://dx.doi.org/10.1038/s42256-021-00302-5},
  DOI = {10.1038/s42256-021-00302-5},
  number = {3},
  journal = {Nature Machine Intelligence},
  publisher = {Springer Science and Business Media LLC},
  author = {Lu,  Lu and Jin,  Pengzhan and Pang,  Guofei and Zhang,  Zhongqiang and Karniadakis,  George Em},
  year = {2021},
  month = {mar},
  pages = {218–229}
}

@article{Raissi2019,
  title = {Physics-informed neural networks: A deep learning framework for solving forward and inverse problems involving nonlinear partial differential equations},
  volume = {378},
  ISSN = {0021-9991},
  url = {http://dx.doi.org/10.1016/j.jcp.2018.10.045},
  DOI = {10.1016/j.jcp.2018.10.045},
  journal = {Journal of Computational Physics},
  publisher = {Elsevier BV},
  author = {Raissi,  M. and Perdikaris,  P. and Karniadakis,  G.E.},
  year = {2019},
  month = {feb},
  pages = {686–707}
}

@article{Perdikaris2017,
  title = {Nonlinear information fusion algorithms for data-efficient multi-fidelity modelling},
  volume = {473},
  ISSN = {1471-2946},
  url = {http://dx.doi.org/10.1098/rspa.2016.0751},
  DOI = {10.1098/rspa.2016.0751},
  number = {2198},
  journal = {Proceedings of the Royal Society A: Mathematical,  Physical and Engineering Sciences},
  publisher = {The Royal Society},
  author = {Perdikaris,  P. and Raissi,  M. and Damianou,  A. and Lawrence,  N. D. and Karniadakis,  G. E.},
  year = {2017},
  month = {feb},
  pages = {20160751}
}

@article{Babaee2016,
  title = {Multi-fidelity modelling of mixed convection based on experimental correlations and numerical simulations},
  volume = {809},
  ISSN = {1469-7645},
  url = {http://dx.doi.org/10.1017/jfm.2016.718},
  DOI = {10.1017/jfm.2016.718},
  journal = {Journal of Fluid Mechanics},
  publisher = {Cambridge University Press (CUP)},
  author = {Babaee,  H. and Perdikaris,  P. and Chryssostomidis,  C. and Karniadakis,  G. E.},
  year = {2016},
  month = {nov},
  pages = {895–917}
}

@article{Pakravan2021,
  title = {Solving inverse-PDE problems with physics-aware neural networks},
  volume = {440},
  ISSN = {0021-9991},
  url = {http://dx.doi.org/10.1016/j.jcp.2021.110414},
  DOI = {10.1016/j.jcp.2021.110414},
  journal = {Journal of Computational Physics},
  publisher = {Elsevier BV},
  author = {Pakravan,  Samira and A. Mistani,  Pouria and Aragon-Calvo,  Miguel A. and Gibou,  Frederic},
  year = {2021},
  month = {sep},
  pages = {110414}
}

@misc{molinaro2023,
      title={Neural Inverse Operators for Solving PDE Inverse Problems}, 
      author={Roberto Molinaro and Yunan Yang and Björn Engquist and Siddhartha Mishra},
      year={2023},
      eprint={2301.11167},
      archivePrefix={arXiv},
      primaryClass={cs.LG},
      url={https://arxiv.org/abs/2301.11167}, 
}

@article{Berg2021,
  title = {Neural networks as smooth priors for inverse problems for PDEs},
  volume = {1},
  ISSN = {2772-4158},
  url = {http://dx.doi.org/10.1016/j.jcmds.2021.100008},
  DOI = {10.1016/j.jcmds.2021.100008},
  journal = {Journal of Computational Mathematics and Data Science},
  publisher = {Elsevier BV},
  author = {Berg,  Jens and Nystr\"{o}m,  Kaj},
  year = {2021},
  month = {sep},
  pages = {100008}
}

@misc{raissi2018ddr,
      title={Multistep Neural Networks for Data-driven Discovery of Nonlinear Dynamical Systems}, 
      author={Maziar Raissi and Paris Perdikaris and George Em Karniadakis},
      year={2018},
      eprint={1801.01236},
      archivePrefix={arXiv},
      primaryClass={math.DS},
      url={https://arxiv.org/abs/1801.01236}, 
}

@article{Pan_2018,
   title={Data-Driven Discovery of Closure Models},
   volume={17},
   ISSN={1536-0040},
   url={http://dx.doi.org/10.1137/18M1177263},
   DOI={10.1137/18m1177263},
   number={4},
   journal={SIAM Journal on Applied Dynamical Systems},
   publisher={Society for Industrial & Applied Mathematics (SIAM)},
   author={Pan, Shaowu and Duraisamy, Karthik},
   year={2018},
   month={jan}, 
pages={2381–2413} }

@article{Rai2000,
  title = {Aerodynamic Design Using Neural Networks},
  volume = {38},
  ISSN = {1533-385X},
  url = {http://dx.doi.org/10.2514/2.938},
  DOI = {10.2514/2.938},
  number = {1},
  journal = {AIAA Journal},
  publisher = {American Institute of Aeronautics and Astronautics (AIAA)},
  author = {Rai,  Man Mohan and Madavan,  Nateri K.},
  year = {2000},
  month = {jan},
  pages = {173–182}
}

@article{Du2021,
  title = {Rapid airfoil design optimization via neural networks-based parameterization and surrogate modeling},
  volume = {113},
  ISSN = {1270-9638},
  url = {http://dx.doi.org/10.1016/j.ast.2021.106701},
  DOI = {10.1016/j.ast.2021.106701},
  journal = {Aerospace Science and Technology},
  publisher = {Elsevier BV},
  author = {Du,  Xiaosong and He,  Ping and Martins,  Joaquim R.R.A.},
  year = {2021},
  month = {jun},
  pages = {106701}
}

@article{Howard2023,
  title = {Multifidelity deep operator networks for data-driven and physics-informed problems},
  volume = {493},
  ISSN = {0021-9991},
  url = {http://dx.doi.org/10.1016/j.jcp.2023.112462},
  DOI = {10.1016/j.jcp.2023.112462},
  journal = {Journal of Computational Physics},
  publisher = {Elsevier BV},
  author = {Howard,  Amanda A. and Perego,  Mauro and Karniadakis,  George Em and Stinis,  Panos},
  year = {2023},
  month = {nov},
  pages = {112462}
}

@article{Lu2022,
  title = {A comprehensive and fair comparison of two neural operators (with practical extensions) based on FAIR data},
  volume = {393},
  ISSN = {0045-7825},
  url = {http://dx.doi.org/10.1016/j.cma.2022.114778},
  DOI = {10.1016/j.cma.2022.114778},
  journal = {Computer Methods in Applied Mechanics and Engineering},
  publisher = {Elsevier BV},
  author = {Lu,  Lu and Meng,  Xuhui and Cai,  Shengze and Mao,  Zhiping and Goswami,  Somdatta and Zhang,  Zhongqiang and Karniadakis,  George Em},
  year = {2022},
  month = {apr},
  pages = {114778}
}

@misc{kohl2024autoreg,
      title={Benchmarking Autoregressive Conditional Diffusion Models for Turbulent Flow Simulation}, 
      author={Georg Kohl and Li-Wei Chen and Nils Thuerey},
      year={2024},
      eprint={2309.01745},
      archivePrefix={arXiv},
      primaryClass={cs.LG},
      url={https://arxiv.org/abs/2309.01745}
}

@inproceedings{lippe2023pderefiner,
  title        = {{PDE-Refiner: Achieving Accurate Long Rollouts with Temporal Neural PDE Solvers}},
  author       = {Phillip Lippe and Bastiaan S. Veeling and Paris Perdikaris and Richard E Turner and Johannes Brandstetter},
  year         = {2023},
  booktitle    = {Thirty-seventh Conference on Neural Information Processing Systems},
  url          = {https://openreview.net/forum?id=Qv6468llWS},
  pages = {1}
}

@misc{ZakariaSmoke2024,
  doi = {10.5281/ZENODO.15629890},
  url = {https://zenodo.org/doi/10.5281/zenodo.15629890},
  author = {Zakaria,  Kemas and Sinisuka,  Anthony Nathan and Palar,  Pramudita and Zuhal,  Lavi},
  title = {Prediction of Smoke-Inflow using Recursive Fourier Neural Operator Dataset},
  publisher = {Zenodo},
  year = {2025},
  copyright = {Creative Commons Attribution 4.0 International},
}

@inproceedings{holl2024phiflow,
  title={${\Phi}_{\text{Flow}}$ ({PhiFlow}): Differentiable Simulations for PyTorch, TensorFlow and Jax},
  author={Holl, Philipp and Thuerey, Nils},
  booktitle={International Conference on Machine Learning},
  year={2024},
  organization={PMLR}
}





\end{document}